\documentclass[journal=jctcce,manuscript=article]{achemso}

\usepackage[version=3]{mhchem} 
\usepackage{graphicx}
\usepackage{amsmath}
\usepackage{amsfonts}
\usepackage{booktabs}
\usepackage{color}
\usepackage{soul}
\usepackage[english]{babel}
\usepackage{multirow}

\newcommand{\vc}[1]{{\bf #1}}
\newcommand{\ket}[1]{\left| #1 \right\rangle}
\newcommand{\bra}[1]{\left\langle #1 \right|}
\newcommand{\braket}[2]{\left\langle #1 \right.\left| #2 \right\rangle}

\newcommand{\onlinecite}[1]{\hspace{-1 ex} \nocite{#1}\citenum{#1}} 

\author{Marco Govoni}
\affiliation{Institute for Molecular Engineering, The University of Chicago, Chicago IL, USA}
\alsoaffiliation{Materials Science Division, Argonne National Laboratory, Argonne IL, USA}
\email{mgovoni@uchicago.edu}
\author{Giulia Galli}
\email{gagalli@uchicago.edu}
\affiliation{Institute for Molecular Engineering, The University of Chicago, Chicago IL, USA}
\alsoaffiliation{Materials Science Division, Argonne National Laboratory, Argonne IL, USA}

\title{Large scale GW calculations}

\abbreviations{MG,GG}
\keywords{GW, quasiparticle, electronic structure}
\date{\today}

\begin{document}

%
%
%
%
%

\begin{abstract}
{We present GW calculations of molecules, ordered and disordered solids and interfaces, which employ an efficient contour deformation technique for frequency integration, and do not require the explicit evaluation of virtual electronic states, nor the inversion of dielectric matrices. We also present a parallel implementation of the algorithm which takes advantage of separable expressions of both the single particle Green's function and the screened Coulomb interaction. The method can be used starting from density functional theory calculations performed with semi-local or hybrid functionals. We applied the newly developed technique to GW calculations of systems of unprecedented size, including water/semiconductor interfaces with thousands of electrons.}
\end{abstract}

\section{Introduction}
\label{sec:introduction}

The accurate description of the excited state properties of electrons plays an important role in many fields of chemistry, physics, and materials science\cite{onida}. For example, the interpretation and prediction of photoemission and opto-electronic spectra of molecules and solids rely on the ability to compute transitions between occupied and virtual electronic states from first principles, as well as their lifetimes\cite{ping_2013_chemrev}. \\
In particular, in the growing field of materials for energy conversion processes -- including solar energy conversion by the photovoltaic effect and solar to fuel generation by water photocatalysis -- it has become key to develop predictive tools to investigate the excited state properties of nanostructures and nanocomposites and of complex interfaces\cite{govoni_2012_CM,marri_2014_CM,wippermann_2014_gwzns}. The latter include solid/solid and solid/liquid interfaces, e.g. between a semiconductor or insulator and water or an electrolyte\cite{Cheng_2010_tio2,wu_2011_pec,cheng_2012_Sprik,Chen_2012_PEC,Anh_2013_si3n4}.\\
In the last three decades, Density Functional Theory (DFT) has been widely used to compute structural and electronic properties of solids and molecules\cite{hohenberg,kohn_sham,dreizler_book,vignale_book,martin_book}. Although successful in describing ground state and thermodynamic properties, and in many ab initio molecular dynamics studies\cite{cui_2011_waterpbe0,gaiduk_2014_pbe0h2onacl}, DFT with both semi-local and hybrid functionals has failed to accurately describe excited state properties of several materials\cite{Marx_book}. However, hybrid functionals have brought great improvement for properties computed with semi-local ones, e.g. for defects in semiconductor and oxides\cite{Alkauskas_2011_hybriddefect,Chen_2013_defecthybrid,Weston_2014_stronzium,Freysoldt_2014_defects}. In particular hybrid functionals with admixing parameters computed self-consistently have shown good performance in reproducing experimental band gaps and dielectric constants of broad classes of systems\cite{Skone_2014_schybrid}. In the case of the electronic properties of surfaces, interfaces (and hence nanostructures), the use of hybrid functionals has in many instances not been satisfactory. Indeed calculations with hybrid functionals yield results for electronic levels that often depend on the mixing parameter chosen for the Hartree-Fock exchange; such parameter is system dependent and there is no known functional yielding satisfactory results for the electronic properties of interfaces composed of materials with substantially different dielectric properties, as different as those of, e.g. water ($\epsilon_\infty$ = 1.78)\cite{epsilon_of_water_exp} and Si ($\epsilon_\infty$ = 11.9)\cite{cardona_book} or water and transition metal oxides of interest for photoelectrodes ($\epsilon_\infty$ = 5-7)\cite{pacchioni_2014_cata}.\\
The use of many body perturbation (MBPT) starting from DFT single particle states has proven accurate for several classes of systems\cite{Hybertsen_1985_qp,hybertsen_qp,godby_qp,farid_1988_gw,engel_1991_gw,Shirley_1993_gw,rohlfing_qp,rojas_1995_gw,rohlfing_1995_gw,aulbur} and it appears to be a promising framework to describe complex nanocomposites and heterogeneous interfaces. MBPT for the calculations of photoemission spectra in the GW approximation\cite{gattibruneval_book_gw}, and of optical spectra by solving approximate forms of the Bethe Salpeter Equation (BSE)\cite{rohlfing_2000_bse} is in principle of general applicability; however its use has been greatly limited by computational difficulties in solving the Dyson's equation and the BSE for realistic systems.\\
Recently we proposed a method to compute quasi particle energies within the $G_0W_0$ approximation (i.e. the non-selfconsistent $GW$ approximation) that does not require the explicit calculation of virtual electronic states, nor the inversion of large dielectric matrices\cite{Viet_2012_GW,Anh_2013_GW}. In addition the method does not use plasmon pole models but instead frequency integrations are explicitly performed and there is one single parameter that controls the accuracy of the computed energies, i.e. the number of eigenvectors and eigenvalues used in the spectral decomposition of the dielectric matrix at zero frequency. The method was successfully used to compute the electronic properties of water\cite{Anh_2014_waterGW} and aqueous solutions\cite{Anhsprik_2014_gwhydroxide} and of heterogeneous solids\cite{wippermann_2014_gwzns}, including crystalline and amorphous samples\cite{Anh_2013_GW}.\\
However the original method contained some numerical approximations in the calculations of the head and wings of the polarizability matrix; most importantly the correlation self-energy was computed on the imaginary axis and obtained in real space by analytic continuation. Finally, although exhibiting excellent scalability, the method was not yet applied to systems with thousands of electrons, e.g. to realistic interfaces, due to the lack of parallelization in its original implementation.\\
In this paper we solved all of the problems listed above, by (i) eliminating numerical approximations in the calculation of the polarizability; (ii) avoiding the use of an analytic continuation and using efficient contour deformation techniques; (iii) providing a parallel implementation of the algorithm based on separable forms of both the single particle Green function and the screened Coulomb interaction. The method presented here can be used starting from DFT orbitals and energies obtained both with semi-local and hybrid functionals. We applied our technique to the calculation of the electronic properties of systems of unprecedented size, including water/semiconductor interfaces with more than one thousand electrons. These calculations allow one to accurately study states at heterogeneous interfaces and to define an electronic thickness of solid/liquid interfaces using MBPT.\\
The rest of the paper is organized as follows. Sec.~\ref{sec:methodintro} describes the $G_0W_0$ methodology that we implemented in a computational package called \texttt{West}. Sec.~\ref{sec:validation} presents results for the ionization potentials of closed and open shell molecules and for the electronic structure of crystalline, amorphous and liquid systems, aimed at verifying and validating the algorithm and code \texttt{West} against previous calculations and measurements. Sec.~\ref{sec:largesystems} presents the study of the electronic properties of finite and extended large systems, i.e. nanocrystals and solid/liquid interfaces, of interest to photovoltaic and photocatalysis applications, respectively. Our conclusions are reported in Sec.~\ref{sec:conclusions}.

\clearpage

\section{Method}
\label{sec:methodintro}

\begin{figure}
\includegraphics[width=0.70\textwidth]{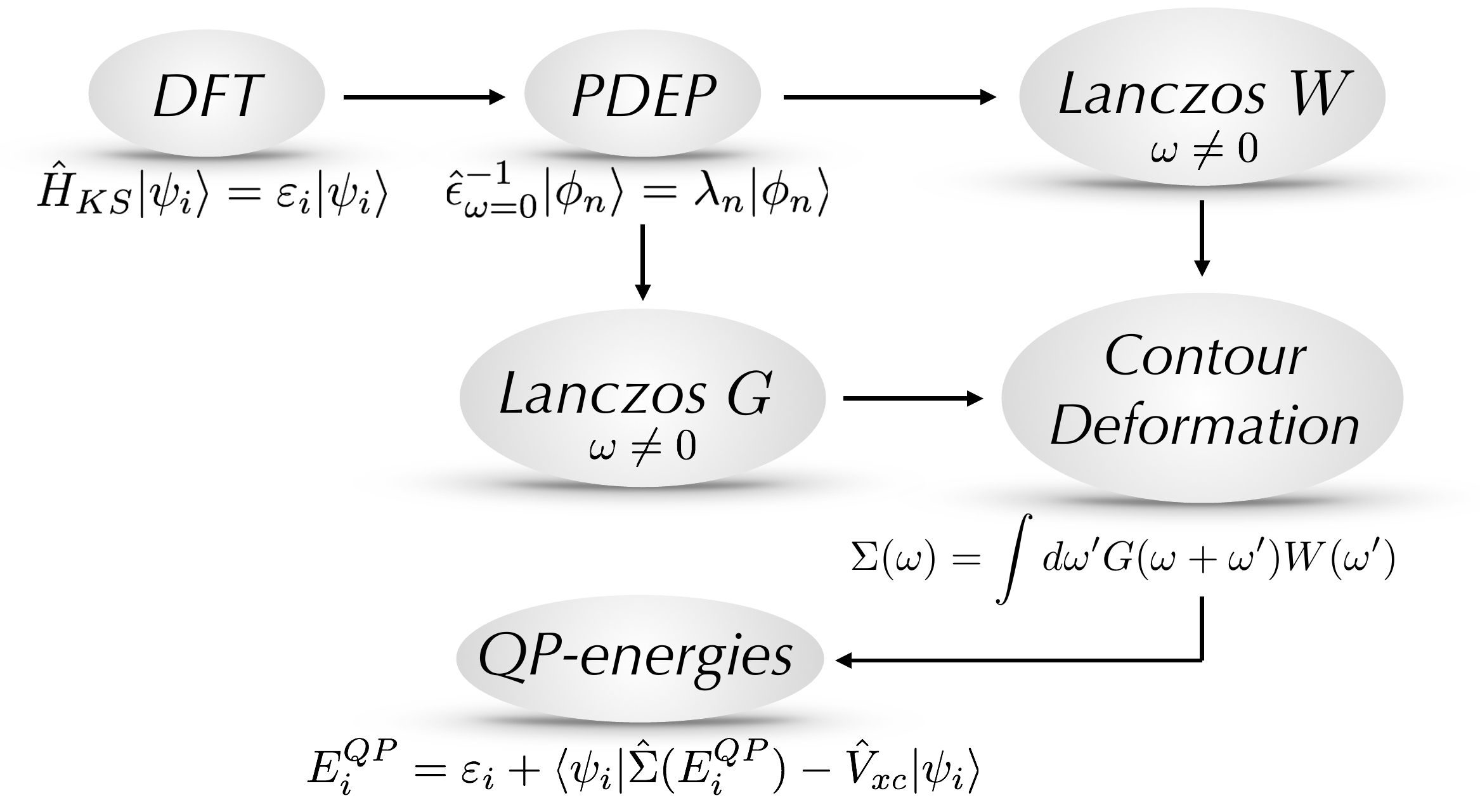}
\caption{(Color online) Schematic representation of the steps involved in the calculations of quasiparticle (QP) energies, within the $G_0W_0$ approximation, using the method proposed in this work. The KS energies ($\epsilon_i$) and occupied orbitals ($\psi_i$) computed at the DFT level are input to the PDEP algorithm, which is used to iteratively diagonalize the static dielectric matrix ($\epsilon^{-1}$) at zero frequency. The set of eigenvectors $\left\{\phi_n\right\}$ constitutes the basis set used to compute both $G$ and $W$ at finite frequencies with the Lanczos algorithm. The frequency integration of Eq.~\eqref{eq:sigmageneral} is carried out using the contour deformation technique. The frequency dependent matrix elements of the electron self-energy are thus obtained and introduced in Eq.~\eqref{eq:QPenergies} to compute the QP energies $E_i^{QP}$.}
  \label{fig:scheme_overview}
\end{figure}

Within DFT, the $n$-th single particle state $\psi_{n\vc{k}\sigma}$ and energy $\varepsilon_{n\vc{k}\sigma}$ of a system of interacting electrons is obtained by solving the Kohn-Sham (KS) equation\cite{hohenberg,kohn_sham,dreizler_book,vignale_book,martin_book}:
\begin{equation}
\hat{H}_{KS}^{\sigma} \left| \psi_{n\vc{k}\sigma} \right\rangle = \varepsilon_{n\vc{k}\sigma} \left| \psi_{n\vc{k}\sigma} \right\rangle \label{eq:secEqforHKS}
\end{equation}
where $\hat{H}_{KS}^{\sigma}=\hat{T}+ \hat{V}_{ion} + \hat{V}_{H} + \hat{V}^{\sigma}_{xc}$ is the KS Hamiltonian, $\hat{T}$ is the kinetic energy operator, and $\hat{V}_{ion}$, $\hat{V}_H$ and $\hat{V}^{\sigma}_{xc}$ are the ionic, Hartree, and exchange-correlation potential operators, respectively. The indexes $\vc{k}$ and $\sigma$ label a wavevector within the first Brillouin zone (BZ) and spin polarization, respectively. Here we consider collinear spin, i.e. decoupled up and down spins.\\
In a fashion similar to Eq.~\eqref{eq:secEqforHKS} one may obtain quasiparticle (QP) states $\psi^{QP}_{n\vc{k}\sigma}$ and QP energies $E^{QP}_{n\vc{k}\sigma}$ by solving the equation:
\begin{equation}
\hat{H}_{QP}^{\sigma} \left| \psi^{QP}_{n\vc{k}\sigma} \right\rangle = E^{QP}_{n\vc{k}\sigma} \left| \psi^{QP}_{n\vc{k}\sigma} \right\rangle \label{eq:secEqforHQP}
\end{equation}
where the QP Hamiltonian $\hat{H}_{QP}^{\sigma}$ is formally obtained by replacing, in Eq.~\eqref{eq:secEqforHKS}, the exchange-correlation potential operator with the electron self-energy operator $\Sigma^\sigma = i G^\sigma W \Gamma $; $G^\sigma$ is the interacting one-particle Green's function, $W$ is the screened Coulomb interaction and $\Gamma$ is the vertex operator\cite{hybertsen_qp,gross_book}. All quantities entering the definition of the self-energy are interdependent and can be obtained self-consistently adopting the scheme suggested by L. Hedin\cite{hedin,hedin_1970_gw,strinati}. In the $GW$ approximation, $\Gamma$ is set equal to the identity, which yields the following expression for the electron self-energy\cite{Aryasetiawan_1998_gw}:
\begin{equation}
\Sigma^\sigma(\vc{r},\vc{r^\prime};\omega) = i \int\limits_{-\infty}^{+\infty} \frac{d\omega^\prime}{2\pi} G^\sigma(\vc{r},\vc{r^\prime};\omega+\omega^\prime) W_{RPA}(\vc{r};\vc{r^\prime};\omega^\prime)\,,  \label{eq:sigmageneral}
\end{equation}
where $W_{RPA}$ is the screened Coulomb interaction obtained in the random phase approximation (RPA).
Due to the non-locality and the frequency dependence of the electron self-energy, a self-consistent solution of Eq.~\eqref{eq:secEqforHQP} is computationally very demanding also for relatively small systems, containing tens of electrons, and usually one evaluates QP energies $E^{QP}_{n\vc{k}\sigma}$ perturbatively:
\begin{eqnarray}
E^{QP}_{n\vc{k}\sigma} &=& \epsilon_{n\vc{k}\sigma} + \bra{\psi_{n\vc{k}\sigma}} \left( \hat{H}^\sigma_{QP} - \hat{H}^\sigma_{KS} \right) \ket{\psi_{n\vc{k}\sigma}} \label{eq:QPpertexp} \\
&=& \epsilon_{n\vc{k}\sigma} +  \bra{\psi_{n\vc{k}\sigma}} \hat{\Sigma}^\sigma (E^{QP}_{n\vc{k}\sigma}) \ket{\psi_{n\vc{k}\sigma}} - \bra{ \psi_{n\vc{k}\sigma} } \hat{V}_{xc}^\sigma  \ket{\psi_{n\vc{k}\sigma}}\,. \label{eq:QPenergies}
\end{eqnarray}
We note that $E^{QP}_{n\vc{k}\sigma}$ enters both the left and right hand side of Eq.~\eqref{eq:QPenergies}, whose solution is usually obtained recursively, e.g. with root-finding algorithms such as the secant method. The use of Eq.~\eqref{eq:QPenergies} to evaluate QP energies from KS states and of the corresponding KS wavefunctions is known as 
the $G_0W_0$ approximation.\\
Within $G_0W_0$, using the Lehmann's representation, the Green's function is:
\begin{equation}
G^\sigma_{KS} (\vc{r},\vc{r^\prime};\omega) = - \sum_{n} \int\limits_{BZ} \frac{d\vc{k}}{(2\pi)^3} \frac{\psi_{n\vc{k}\sigma}(\vc{r})\psi^\ast_{n\vc{k}\sigma}(\vc{r^\prime})}{ \epsilon_{n\vc{k}\sigma} - \omega - i \eta \text{sign}( \varepsilon_{n\vc{k}\sigma} - \varepsilon_F)} \label{eq:Glehmann}
\end{equation}
where $\eta$ is a small positive quantity and $\varepsilon_F$ is the Fermi energy. In Eq.~\eqref{eq:Glehmann} we used the subscript KS to indicate that the Green's function is evaluated using the KS orbitals obtained by solving Eq.~\eqref{eq:secEqforHKS}.\\
In Ref.~[\onlinecite{Viet_2012_GW,Anh_2013_GW}] an algorithm was introduced to compute the self-energy matrix elements of Eq.~\eqref{eq:QPenergies} without the need to evaluate explicitly empty (virtual) electronic states, by using a technique called projective eigendecomposition of the dielectric screening (PDEP). A diagram of the method is reported in Fig.~\ref{fig:scheme_overview}. After KS single particle orbitals and energies are obtained using semilocal or hybrid functionals, the screened Coulomb interaction is computed using a basis set built from the eigenpotentials of the static dielectric matrix at zero frequency. In this way $W_{RPA}$ entering Eq.~\eqref{eq:sigmageneral} is expressed in a separable form, similar to that of $G^\sigma_{KS}$ in Eq.~\eqref{eq:Glehmann}. In the following sections we describe in detail all the steps outlined in Fig.~\ref{fig:scheme_overview}. The separable form of $W_{RPA}$ is given in Sec.~\ref{sec:pdep}. Calculation of the polarizability and the spectral decomposition of the dielectric matrix are described in Sec.~\ref{sec:fromchi0tochi} and~\ref{sec:pdepmethod}, respectively. Matrix elements of $G_{KS}$ and $W_{RPA}$ are then obtained without the explicit use of empty electronic states and simultaneously at several frequencies by using a deflated Lanczos technique, described in Sec.~\ref{sec:lglw}. Finally the frequency integration is carried out by introducing a contour deformation method, as described in Sec.~\ref{sec:contourdef}. The use of the analytic continuation used in the original method of Ref.~[\onlinecite{Viet_2012_GW,Anh_2013_GW}] is thus avoided.

%
%


\subsection{Separable form of the screened Coulomb interaction}
\label{sec:pdep}
In order to solve equation Eq.~\eqref{eq:QPenergies} and obtain QP energies, one needs to compute the matrix elements of the electron self-energy between KS states, which in the $G_0W_0$ approximation is given by Eq.~\eqref{eq:sigmageneral}. The Green's function may be expressed in a fully separable form using its Lehmann's representation, Eq.~\eqref{eq:Glehmann}.
However $W_{RPA}$ is not trivially separable; it is given as the sum of two terms:
\begin{equation}
W_{RPA}(\vc{r},\vc{r^\prime};\omega) = v(\vc{r},\vc{r^\prime}) + W_{p}(\vc{r},\vc{r^\prime};\omega) \label{eq:Wrparealspace}
\end{equation}
where $v(\vc{r},\vc{r^\prime}) = \frac{e^2}{\left|\vc{r}-\vc{r^\prime}\right|}$ is the bare Coulomb interaction\bibnote{We use everywhere atomic Rydberg units.} and $W_p$ is a nonlocal and frequency dependent function. Using Eq.~\eqref{eq:Wrparealspace}, we write the self-energy as the sum of two contributions $\Sigma^\sigma = \Sigma^\sigma_X + \Sigma^\sigma_C$, where only the latter depends on the frequency:
\begin{eqnarray}
\Sigma^\sigma_X(\vc{r},\vc{r^\prime}) &=& i \int\limits_{-\infty}^{+\infty} \frac{d\omega^\prime}{2\pi} G^\sigma_{KS}(\vc{r},\vc{r^\prime};\omega+\omega^\prime) v(\vc{r};\vc{r^\prime})  \label{eq:sigmax} \\
& = & - 
\sum_{n=1}^{N_{occ}^\sigma} \int\limits_{BZ} \frac{d\vc{k}}{(2\pi)^3}  \psi_{n\vc{k}\sigma}(\vc{r}) v(\vc{r},\vc{r^\prime}) \psi^\ast_{n\vc{k}\sigma}(\vc{r^\prime}) \label{eq:fock}
\end{eqnarray}
$N_{occ}^\sigma$ is the number of occupied states with spin $\sigma$; $\Sigma^\sigma_X$ is usually called exchange self-energy because it is formally equivalent to the Fock exact exchange operator\bibnote{The calculation of the Fock exact exchange matrix elements in reciprocal space is done employing the Gygi-Baldereschi method.\cite{gygibaldereschi_1986_sing}}; 
\begin{equation}
\Sigma^\sigma_C(\vc{r},\vc{r^\prime};\omega) = i \int\limits_{-\infty}^{+\infty} \frac{d\omega^\prime}{2\pi} G^\sigma_{KS}(\vc{r},\vc{r^\prime};\omega+\omega^\prime) W_p(\vc{r},\vc{r^\prime};\omega^\prime) \label{eq:sigmac}
\end{equation}
$\Sigma^\sigma_C$ is referred to as correlation self-energy. Using Eq.s~\eqref{eq:sigmax}-\eqref{eq:sigmac} the QP Hamiltonian of Eq.~\eqref{eq:secEqforHQP} may be expressed as:
\begin{equation}
\hat{H}_{QP}(\omega) = \hat{T} + \hat{V}_{ion} + \hat{V}_{HF}^\sigma + \hat{\Sigma}_C^\sigma(\omega)
\end{equation}
where $\hat{V}_{HF}^\sigma$ is the Hartree-Fock potential operator. The ionic potential $\hat{V}_{ion}$ is treated within the pseudopotential approach\bibnote{Although the presented methodology is general, for the purpose of this work we have considered only normconserving pseudopotentials\cite{focher_1991_pseudoisphysicalion}.}.\\
In this work we express $W_p$ in a separable form by adopting the projective dielectric eigendecomposition (PDEP) technique, proposed in Ref.~[\onlinecite{Wilson_2008_PDEP}-\onlinecite{Wilson_2009_PDEP}], and we use a plane wave basis set to express the single particle wave functions and charge density, within periodic boundary conditions:
\begin{equation} 
\psi_{n\vc{k}\sigma}(\vc{r}) = e^{i\vc{k}\cdot\vc{r}} u_{n\vc{k}\sigma}(\vc{r}) = \sum_{\vc{G}} c_{n\vc{k}\sigma}(\vc{G})e^{i(\vc{k+G})\cdot\vc{r}} \label{eq:planewaveexpansionofwfc}
\end{equation}
where $\vc{G}$ is a reciprocal lattice vector, $c_{n\vc{k}\sigma}(\vc{G}) = \frac{1}{\Omega} \int_{\Omega} d\vc{r}\, u_{n\vc{k}\sigma}(\vc{r}) e^{-i \vc{G}\cdot\vc{r}}$ and $\Omega$ is the unit cell volume. 
In Eq.~\eqref{eq:planewaveexpansionofwfc} all reciprocal lattice vectors such that $\frac{1}{2}\left|\vc{k+G}\right|^2<E_{cutwfc}$ are included in the summation.
Using a plane wave description also for $W_p$ we have 
\begin{equation}
W_{RPA}(\vc{r},\vc{r^\prime};\omega) = 
\int\limits_{BZ}\frac{d\vc{q}}{(2\pi)^3} \sum_{\vc{G}\vc{G^\prime}} e^{i(\vc{q+G})\cdot\vc{r}} \left[ v_{\vc{G}\vc{G^\prime}} + W^p_{\vc{G}\vc{G^\prime}}(\vc{q};\omega) \right] e^{-i(\vc{q+G^\prime})\cdot\vc{r^\prime}}   
\end{equation}
where $v_{\vc{G}\vc{G^\prime}} = \frac{4\pi e^2}{|\vc{q+G}|^2} \delta_{\vc{G}\vc{G^\prime}}$ ($\delta$ is the Kronecker delta) and 
\begin{equation}
W^p_{\vc{G}\vc{G^\prime}}(\vc{q};\omega)  = \frac{\bar{\chi}_{\vc{G}\vc{G^\prime}} (\vc{q};\omega)}{|\vc{q+G}||\vc{q+G^\prime}|} \,. \label{eq:Wpnondiag} 
\end{equation}
In Eq.~\eqref{eq:Wpnondiag} we have introduced the symmetrized reducible polarizability $\bar{\chi}$, related to the symmetrized inverse dielectric matrix $\bar{\epsilon}^{-1}$ by the relation:
\begin{equation}
\bar{\epsilon}^{-1}_{\vc{G}\vc{G^\prime}} (\vc{q};\omega) = \delta_{\vc{G}\vc{G^\prime}} + \bar{\chi}_{\vc{G}\vc{G^\prime}} (\vc{q};\omega) \,. \label{eq:epsilonandchi}
\end{equation}
The symmetrized form $\bar{\chi}$ of the polarizability $\chi$ is
\begin{equation}
\bar{\chi}_{\vc{G}\vc{G^\prime}} = \frac{\sqrt{4\pi e^2}}{|\vc{q+G}|} \chi_{\vc{G}\vc{G^\prime}} \frac{\sqrt{4\pi e^2}}{|\vc{q+G^\prime}|} \,\,. \label{eq:symmoperator}
\end{equation}
The reducible polarizability $\chi$ is related to the irreducible polarizability $\chi_0$ by the Dyson's equation, which within the RPA reads:
\begin{equation}
\chi_{\vc{G}\vc{G^\prime}} = \chi^0_{\vc{G}\vc{G^\prime}} + \sum_{\vc{G_1},\vc{G_2}}\chi^0_{\vc{G}\vc{G_1}} v_{\vc{G_1}\vc{G_2}} \chi_{\vc{G_2}\vc{G^\prime}} 
\end{equation}
or in terms of symmetrized polarizabilites:
\begin{equation}
\bar{\chi} = ( 1 - \bar{\chi}^0 )^{-1} \bar{\chi}^0 \,.  \label{eq:linearalgebrachichi0}
\end{equation}
Within a plane wave representation each quantity in Eq.~\eqref{eq:linearalgebrachichi0} is a matrix of dimension $N_{pw}^2$, and in principle $\bar{\chi}$ can be obtained from $\bar{\chi}^0$ via simple linear algebra operations. 
In practice, the matrices of Eq.~\eqref{eq:linearalgebrachichi0} may contain millions of rows and columns for realistic systems; for example for a silicon nanocrystal with 465 atoms, placed in a cubic box of edge $90\,\text{bohr}$, 1.5 million plane waves are needed in the expansion of Eq.~\eqref{eq:planewaveexpansionofwfc} with $E_{cutwfc}=25\,\text{Ry}$. 
It is thus important to find alternative representations of $\bar{\chi}$ and reduce the number of elements to compute. One could think of a straightforward spectral decomposition:
\begin{equation}
\bar{\chi}_{\vc{G}\vc{G^\prime}} (\vc{q};\omega) = \sum_{i=1}^{N_{pdep}} \phi_i\left(\vc{q+G};\omega\right) \lambda_{i}\left(\vc{q};\omega\right) \phi^\ast_j\left(\vc{q+G};\omega\right) \label{eq:eigendecompanyfreq}
\end{equation}
where $\phi_i$ and $\lambda_i$ are the eigenvectors and eignvalues of $\bar{\chi}$, respectively. Unfortunately this strategy is still too demanding from a computational standpoint, as it implies finding eigenvectors and eigenvalues at multiple frequencies. \\
A computationally more tractable representation may be obtained using the spectral decomposition of $\bar{\chi}_0$ at $\omega=0$. As apparent from Eq.~\eqref{eq:linearalgebrachichi0}, eigenvectors of $\bar{\chi}$ are also eigenvectors of $\bar{\chi}^0$; the latter is easier to iteratively diagonalize than $\bar{\chi}$, and the frequency dependency may be dealt with iterative techniques, starting from the solution at $\omega =0$, as discussed in Sec.~\ref{sec:lglw}. Therefore we proceed by solving the secular equation for $\bar{\chi}^0$ only at $\omega=0$, generating what we call the PDEP basis set $\left\{ \left| \phi_i\right\rangle \; : \; i=1,N_{pdep}\right\}$, which is used throughout this work to represent the polarizability $\bar{\chi}$:
\begin{equation}
\bar{\chi}_{\vc{G}\vc{G^\prime}} (\vc{q};\omega) = \sum_{\substack{i=1,\\j=1}}^{N_{pdep}} \phi_i\left(\vc{q+G}\right) \Lambda_{ij}\left(\vc{q};\omega\right) \phi^\ast_j\left(\vc{q+G}\right); \label{eq:eigendecompzerofreq}
\end{equation}
here $\Lambda_{ij}\left(\vc{q};\omega\right)$ is a matrix of dimension $N_{pdep}^2$. In general $N_{pdep} \ll N_{pw}$,\cite{Wilson_2008_PDEP,Wilson_2009_PDEP} leading to substantial computational savings.\bibnote{The approach presented here scales as the fourth power of the system size\cite{Anh_2013_GW}, as conventional approaches do. However the computational workload of our method represents a substantial improvement over that of conventional approaches, because of a much more favorable pre-factor: the scaling is $N^2_{occ}\times N_{pdep} \times N_{pw}$ instead of $N_{occ}\times N_{empt} \times N_{pw}^2$, where $N_{occ}$ ($N_{empt}$) is the number of occupied (empty) states, $N_{pw}$ in the number of plane waves and $N_{pdep}$ is the number of eigenpotentials used in the PDEP expansion of the static dielectric screening.} The $N_{pdep}$ functions $\phi_i$ may be computed by solving the Sternheimer equation\cite{sternheimer_1954}, without explicitly evaluating empty (virtual) electronic states. In addition, $N_{pdep}$ turns out to be the only parameter that controls the accuracy of the expansion in Eq.~\eqref{eq:eigendecompzerofreq}. The details of the derivation of the PDEP basis set are given in Sec.~\ref{sec:pdepmethod}. We note that alternative basis sets, based on the concepts related to maximally localized Wannier functions, have been proposed in the literature to reduce the dimensionality of the polarizability matrices\cite{Umari_2009_GWW}.\\
By defining $\tilde{\phi}_i \left(\vc{q+G}\right)  = \frac{\phi_i \left(\vc{q+G}\right)}{|\vc{q+G}| } $, we formally obtain the desired separable form for $W_p$:
\begin{equation}
W_p(\vc{r},\vc{r^\prime};\omega) = 
\int\limits_{BZ}\frac{d\vc{q}}{(2\pi)^3} 
\sum_{\substack{i=1,\\j=1}}^{N_{pdep}}
\tilde{\phi}_i \left(\vc{q};\vc{r}\right) 
\Lambda_{ij}\left(\vc{q};\omega\right)
\tilde{\phi}^\ast_j \left(\vc{q};\vc{r^\prime}\right). \label{eq:Wseparablecomplete}
\end{equation}
The scaling operation used to define $\tilde{\phi}_i$ is divergent in the long wavelength limit ($\vc{q}\to\vc{0}$) and for $\vc{G}=\vc{0}$. However such divergence can be integrated yielding:
\begin{equation}
W_p(\vc{r},\vc{r^\prime};\omega) = \Xi(\omega)  + \frac{1}{\Omega}  
\sum_{\substack{i=1,\\j=1}}^{N_{pdep}} 
\tilde{\phi}_i \left(\vc{r}\right) 
\Lambda_{ij}(\omega) 
\tilde{\phi}^\ast_j \left(\vc{r^\prime}\right) \,,
\label{eq:Wchihb}
\end{equation}
where 
\begin{equation}
\Xi(\omega) =  4\pi e^2 \int\limits_{R_{\vc{q}=0}} \frac{d\vc{q}}{(2\pi)^3} \frac{\bar{\chi}_{\vc{0}\vc{0}} (\vc{q};\omega)}{q^2}\,. \label{eq:integrationxiq}
\end{equation}
In Eq.~\eqref{eq:integrationxiq} the integration is evaluated on the region $R_{\vc{q}=0}$ of the $BZ$ enclosing the $\Gamma$-point (i.e. $\vc{q}=\vc{0}$).\bibnote{The divergence of the Coulomb potential present in Eq.~\eqref{eq:integrationxiq} can be numerically removed using spherical coordinates. The specific shape of the BZ is taken into account by using a Monte Carlo integration method.} \\
In the $\vc{q}\to \vc{0}$ limit, we can now write the matrix elements of $\Sigma_C$ using: i) the separable form of $W_p$ of Eq.~\eqref{eq:Wchihb} and ii) the expression of $G_{KS}$, given in Eq.~\eqref{eq:Glehmann}, in terms of projector operators:
\begin{equation}
\hat{G}^\sigma_{KS} (\omega) = \int\limits_{BZ} \frac{d\vc{k}}{(2\pi)^3} 
\hat{P}^{\vc{k}\sigma}_{v} \hat{O}^\sigma_{KS} \left( \omega -i\eta  \right) \hat{P}^{\vc{k}\sigma}_{v} + 
\int\limits_{BZ} \frac{d\vc{k}}{(2\pi)^3} 
\hat{P}^{\vc{k}\sigma}_{c} \hat{O}^\sigma_{KS} \left( \omega +i\eta  \right) \hat{P}^{\vc{k}\sigma}_{c}
\end{equation}
where 
\begin{equation}
\hat{O}^\sigma_{KS} \left( \omega \right) = -\left(\hat{H}^\sigma_{KS} - \omega\right)^{-1}\,, \label{eq:Odef}
\end{equation}
$\hat{P}^{\vc{k}\sigma}_{v}=\sum_{n=1}^{N^\sigma_{occ}}\ket{\psi_{n\vc{k}\sigma}}\bra{\psi_{n\vc{k}\sigma}}$ and $\hat{P}^{\vc{k}\sigma}_{c}=\sum_{n=N^\sigma_{occ}+1}^{+\infty}\ket{\psi_{n\vc{k}\sigma}}\bra{\psi_{n\vc{k}\sigma}}$ are the projector operator over the occupied and unoccupied manyfold of states belonging to k-point $\vc{k}$ and spin $\sigma$, respectively\bibnote{In the present formulation we considered integer occupation numbers and only the dielectric response given by interband transitions.}. Hence we have: 
\begin{equation}
\bra{\psi_{n\vc{k}\sigma}} \Sigma^\sigma_C(\omega) \ket{\psi_{n\vc{k}\sigma}} = A_{n\vc{k}\sigma}(\omega) + B_{n\vc{k}\sigma}(\omega) + C_{n\vc{k}\sigma}(\omega) + D_{n\vc{k}\sigma}(\omega) \,, \label{eq:SCABCD}
\end{equation}
where $A_{n\vc{k}\sigma}$ and $C_{n\vc{k}\sigma}$ ($B_{n\vc{k}\sigma}$ and $D_{n\vc{k}\sigma}$) are contributions to the correlation self-energy originating from occupied (empty) states:
\begin{equation}
A_{n\vc{k}\sigma}(\omega)=i \int\limits_{-\infty}^{+\infty} \frac{d\omega^\prime}{2\pi} \Xi(\omega^\prime) \bra{\psi_{n\vc{k}\sigma}} \hat{P}^{\vc{k}\sigma}_{v} \hat{O}^\sigma_{KS}  \left( \omega + \omega^\prime - i\eta \right) \hat{P}^{\vc{k}\sigma}_{v}\ket{\psi_{n\vc{k}\sigma}} \label{eq:SCA}
\end{equation}
\begin{equation}
B_{n\vc{k}\sigma}(\omega)=i \int\limits_{-\infty}^{+\infty} \frac{d\omega^\prime}{2\pi} \Xi(\omega^\prime) \bra{\psi_{n\vc{k}\sigma}} \hat{P}^{\vc{k}\sigma}_{c} \hat{O}^\sigma_{KS} \left( \omega + \omega^\prime + i\eta \right) \hat{P}^{\vc{k}\sigma}_{c} \ket{\psi_{n\vc{k}\sigma}}  \label{eq:SCB}
\end{equation}
\begin{equation}
C_{n\vc{k}\sigma}(\omega)=\frac{i}{\Omega} \int\limits_{-\infty}^{+\infty} \frac{d\omega^\prime}{2\pi} \sum_{\substack{i=1,\\j=1}}^{N_{pdep}} \Lambda_{ij}(\omega^\prime) \bra{\phi^i_{n\vc{k}\sigma}} \hat{P}^{\vc{k}\sigma}_{v} \hat{O}^\sigma_{KS} \left( \omega + \omega^\prime - i\eta \right) \hat{P}^{\vc{k}\sigma}_{v}\ket{\phi^j_{n\vc{k}\sigma}} \label{eq:SCC}
\end{equation}
\begin{equation}
D_{n\vc{k}\sigma}(\omega)=\frac{i}{\Omega} \int\limits_{-\infty}^{+\infty} \frac{d\omega^\prime}{2\pi} \sum_{\substack{i=1,\\j=1}}^{N_{pdep}} \Lambda_{ij}(\omega^\prime) \bra{\phi^i_{n\vc{k}\sigma}} \hat{P}^{\vc{k}\sigma}_{c} \hat{O}^\sigma_{KS} \left( \omega + \omega^\prime + i\eta \right) \hat{P}^{\vc{k}\sigma}_{c} \ket{\phi^j_{n\vc{k}\sigma}} \label{eq:SCD}
\end{equation}
We have defined $\phi^j_{n\vc{k}\sigma}(\vc{r})=\psi_{n\vc{k}\sigma}\left(\vc{r}\right)\tilde{\phi}^\ast_{j}\left(\vc{r}\right)$. The quantities 
$A_{n\vc{k}\sigma}$, $B_{n\vc{k}\sigma}$, $C_{n\vc{k}\sigma}$ and $D_{n\vc{k}\sigma}$ entering Eq.~\eqref{eq:SCABCD} are now in a form where  iterative techniques (see Sec.~\ref{sec:lglw}) can be applied to obtain the matrix elements of the correlation self-energy. Moreover, because of the completeness of energy eigenstates ($\hat{P}^{\vc{k}\sigma}_{c}=1-\hat{P}^{\vc{k}\sigma}_{v}$), we may compute all quantities in Eq.s~\eqref{eq:SCA}-\eqref{eq:SCD} considering only occupied states. The integration over the frequency domain will be discussed in Sec.~\eqref{sec:contourdef}.


\subsection{Polarizability within the random phase approximation}
\label{sec:fromchi0tochi}

Here we discuss how to compute the polarizability $\bar{\chi}$ from $\bar{\chi}^0$ within the RPA, in the long wavelength limit ($\vc{q}\to\vc{0}$), without explicitly evaluating electronic empty states. The Fourier components of the symmetrized irreducible polarizability $\bar{\chi}^0$ are given by the Adler-Wiser expression\cite{adler,wiser}, which contains an explicit summation over unoccupied states:
\begin{eqnarray}
\bar{\chi}^0_{\vc{G}\vc{G^\prime}}\left(\vc{q};\omega\right) &=& - 4\pi e ^2 \sum_\sigma \sum_{n=1}^{N_{occ}^\sigma}\sum_{m=N_{occ}^\sigma+1}^{+\infty} \int\limits_{BZ} \frac{d\vc{k}}{(2\pi)^3} \frac{ \rho^\ast_{mn\vc{k}\sigma} (\vc{q},\vc{G}) \rho_{mn\vc{k}\sigma}(\vc{q},\vc{G^\prime})}{|\vc{q+G}||\vc{q+G^\prime}|}   \times \nonumber\\
&&\times \left[\frac{1}{\epsilon_{m\vc{k}\sigma} - \epsilon_{n\vc{k-q}\sigma} + \omega -i\eta  }+\frac{1}{\epsilon_{m\vc{k}\sigma} - \epsilon_{n\vc{k-q}\sigma} - \omega -i\eta }\right] \label{eq:AdlerWiser}
\end{eqnarray}
where the matrix element
\begin{equation}
\rho_{mn\vc{k}\sigma} (\vc{q},\vc{G}) = 
\bra{\psi_{m\vc{k}\sigma}} 
e^{i(\vc{q+G})\cdot \vc{r}}  
\ket{\psi_{n\vc{k-q}\sigma}} \label{eq:oscillator}
\end{equation}
is often referred to as oscillator strength; it has the following properties:
\begin{eqnarray}
\left. \rho_{mn\vc{k}\sigma} (\vc{q},\vc{G}=\vc{0}) \right|_{\vc{q}\to\vc{0}} &=& \delta_{nm} \label{eq:oscillator0} \\
\left.\nabla_\vc{q} \rho_{mn\vc{k}\sigma} (\vc{q},\vc{G}=\vc{0}) \right|_{\vc{q}\to\vc{0}} &=& i \bra{\psi_{m\vc{k}\sigma}} 
\vc{r} 
\ket{\psi_{n\vc{k}\sigma}}. \label{eq:oscillator1}
\end{eqnarray}
Following Ref.~[\onlinecite{baroni_localfields,hybertsen_dm}], we partition the polarizability of Eq.~\eqref{eq:AdlerWiser} into head ($\vc{G}=\vc{G}^\prime =\vc{0}$), wings ($\vc{G}=\vc{0}$,$\vc{G}^\prime \neq \vc{0}$ or $\vc{G}  \neq \vc{0}$,$\vc{G}^\prime = \vc{0}$) and body ($\vc{G}\neq\vc{0}$ and $\vc{G}^\prime\neq\vc{0}$) elements. The $\vc{q}\to\vc{0}$ limit of the body, which we call $B_{\vc{G}\vc{G^\prime}}$, is analytic, while the limits of the head and wings are non-analytic, i.e. they depend on the Cartesian direction along which the limit is performed. The long wavelength limits of the head, body and wings of the polarizability matrix are summarized in Table~\ref{tbl:limitqchi0}.
%
%
%
\begin{table}
  \begin{tabular}{c|cc}
    \hline\hline
     $\bar{\chi}^0_{\vc{G}\vc{G^\prime}}(\vc{q}\to\vc{0};\omega)$       &  $\vc{G^\prime}=\vc{0}$ & $\vc{G^\prime}\neq\vc{0}$ \\
    \hline
    $\vc{G}=\vc{0}$  & $\sum_{\alpha\beta} q_\alpha F_{\alpha\beta}(\omega) q_{\beta}/q^2$ & $-i \sum_\alpha q_{\alpha} U_{\alpha\vc{G^\prime}}(\omega)/q$ \\
    $\vc{G}\neq\vc{0}$  & $i \sum_\beta U_{\vc{G}\beta}(\omega)q_{\beta}/q$ & $B_{\vc{G}\vc{G^\prime}}$ \\
    \hline\hline
  \end{tabular}
  \caption{The long wavelength limit ($\vc{q}\to\vc{0}$) of the head, wing and body elements of the polarizability $\bar{\chi}^0_{\vc{G}\vc{G^\prime}}(\omega)$ are given in the second and third columns: $U_{\vc{G}\beta}(\omega)=-i4\pi e^2 \frac{\partial}{\partial q_\beta}\chi_{\vc{G}\vc{0}}(\omega)$ and $F_{\alpha\beta}(\omega)=4\pi e^2 \frac{\partial^2}{\partial q_\alpha\partial q_\beta}\chi_{\vc{0}\vc{0}}(\omega)$ are evaluated using Eq.~\eqref{eq:oscillator1} and yield the linear and quadratic terms in the Taylor expansion of $\chi^0(\omega)$ around $\vc{q}=\vc{0}$, respectively.}
\label{tbl:limitqchi0}
\end{table}
Using the PDEP basis set we obtain:
\begin{equation}
U_{\alpha j }(\omega) = \sum_{\vc{G}^\prime} U_{\alpha\vc{G^\prime}} (\omega) \tilde{\phi}_{j}(\vc{G^\prime})
\end{equation}
\begin{equation}
B_{i j }(\omega) = \sum_{\vc{G}\vc{G^\prime}} \tilde{\phi}^\ast_{i}(\vc{G}) B_{\vc{G}\vc{G^\prime}} (\omega) \tilde{\phi}_{j}(\vc{G^\prime})
\end{equation} 
We can now express all the quantities in Table~\ref{tbl:limitqchi0} without any explicit summation over empty (virtual) states:
\begin{equation}
F_{\alpha\beta} (\omega) = 4\pi e ^2 \sum_{\sigma}\sum_{n=1}^{N_{occ}^\sigma} \int\limits_{BZ} \frac{d\vc{k}}{(2\pi)^3} 
\bra{\xi^\alpha_{n\vc{k}\sigma}}  \hat{P}^{\vc{k}\sigma}_{c} 
\left[ \hat{O}^\sigma_{KS}\left( \epsilon_{n\vc{k}\sigma} - \omega +i\eta  \right) + \hat{O}^\sigma_{KS}\left( \epsilon_{n\vc{k}\sigma} + \omega +i\eta  \right) \right]   \hat{P}^{\vc{k}\sigma}_{c} \ket{\xi^\beta_{n\vc{k}\sigma}} \label{eq:finalF}
\end{equation}
\begin{equation}
U_{\alpha j} (\omega) =  4\pi e ^2 \sum_{\sigma}\sum_{n=1}^{N_{occ}^\sigma} \int\limits_{BZ} \frac{d\vc{k}}{(2\pi)^3} 
\bra{\xi^\alpha_{n\vc{k}\sigma}}  \hat{P}^{\vc{k}\sigma}_{c} 
\left[ \hat{O}^\sigma_{KS}\left( \epsilon_{n\vc{k}\sigma} - \omega +i\eta  \right) + \hat{O}^\sigma_{KS}\left( \epsilon_{n\vc{k}\sigma} + \omega +i\eta  \right) \right]    \hat{P}^{\vc{k}\sigma}_{c} \ket{\xi^j_{n\vc{k}\sigma}} \label{eq:finalU1}
\end{equation}
\begin{equation}
U_{i\alpha} (\omega) =  4\pi e ^2 \sum_{\sigma}\sum_{n=1}^{N_{occ}^\sigma} \int\limits_{BZ} \frac{d\vc{k}}{(2\pi)^3} 
\bra{\xi^i_{n\vc{k}\sigma}}  \hat{P}^{\vc{k}\sigma}_{c} 
\left[ \hat{O}^\sigma_{KS}\left( \epsilon_{n\vc{k}\sigma} - \omega +i\eta  \right) + \hat{O}^\sigma_{KS}\left( \epsilon_{n\vc{k}\sigma} + \omega +i\eta  \right) \right]   \hat{P}^{\vc{k}\sigma}_{c}  \ket{\xi^\beta_{n\vc{k}\sigma}} \label{eq:finalU2}
\end{equation}
\begin{equation}
B_{ij} (\omega) = 4\pi e ^2 \sum_{\sigma}\sum_{n=1}^{N_{occ}^\sigma} \int\limits_{BZ} \frac{d\vc{k}}{(2\pi)^3} 
\bra{\xi^i_{n\vc{k}\sigma}}  \hat{P}^{\vc{k}\sigma}_{c} 
\left[ \hat{O}^\sigma_{KS}\left( \epsilon_{n\vc{k}\sigma} - \omega +i\eta  \right) + \hat{O}^\sigma_{KS}\left( \epsilon_{n\vc{k}\sigma} + \omega +i\eta  \right) \right]  \hat{P}^{\vc{k}\sigma}_{c} \ket{\xi^j_{n\vc{k}\sigma}} \label{eq:finalB}
\end{equation}
Note that the greek letters $\alpha$ and $\beta$ identify Cartesian directions, while the roman letters $i$ and $j$ label the eigevectors of $\bar{\chi}^0$ at $\omega=0$, i.e. the elements of the PDEP basis set. We have also defined the auxiliary functions $\xi^i_{n\vc{k}\sigma}(\vc{r}) = \psi_{n\vc{k}\sigma}(\vc{r}) \tilde{\phi}_i(\vc{r})$ and $\xi^\alpha_{n\vc{k}\sigma}(\vc{r}) = \hat{P}_{c}^{\vc{k}\sigma} r_\alpha \ket{\psi_{n\vc{k}\sigma}} $. Within periodic boundary conditions the position operator is ill-defined and $\xi^\alpha_{n\vc{k}\sigma}(\vc{r})$ is obtained by solving the linear system
\begin{equation}
\left(\hat{H}^\sigma_{KS}- \epsilon_{n\vc{k}\sigma} \right) \ket{\xi^\alpha_{n\vc{k}\sigma}} = \hat{P}_{c}^{\vc{k}\sigma} \left[ \hat{H}^\sigma_{KS}, r_\alpha \right] \ket{\psi_{n\vc{k}\sigma}} \label{eq:commu}
\end{equation}
where the commutator of the KS Hamiltonian with the position operator includes the contribution of the non-local part of the pseudopotential\cite{baroni_localfields,hybertsen_dm}. Once $\bar{\chi}^0$ is obtained, $\bar{\chi}$ is computed using Eq.~\eqref{eq:linearalgebrachichi0}. \\
The quantities required to evaluate Eq.~\eqref{eq:SCABCD} are the following\cite{Freysoldt_2007_anisotropy}:
\begin{equation}
\Xi(\omega) =  \frac{1-k(\omega)}{k(\omega)} \int\limits_{R_{\vc{q}=0}} \frac{d\vc{q}}{(2\pi)^3} \frac{4\pi e^2}{|\vc{q}|^2} 
\end{equation}
\begin{equation}
\vc{\Lambda}(\omega) =  [\vc{1}-\vc{B}(\omega)]^{-1} \vc{B}(\omega) +  \frac{1}{k(\omega)} [\vc{1}-\vc{B}(\omega)]^{-1} \boldsymbol\mu (\omega)  [\vc{1}-\vc{B}(\omega)]^{-1} \label{eq:Bfinal}
\end{equation}
with $k(\omega) = 1-f(\omega) - \text{Tr} \left\{ \boldsymbol\mu (\omega)  [\vc{1} - \vc{B}(\omega)]^{-1} \right\}$; $f(\omega) = \frac{1}{3} \sum_\alpha F_{\alpha\alpha}(\omega)$ and the matrix elements of $\mu_{ij}(\omega) = \frac{1}{3} \sum_\alpha U_{i\alpha}(\omega)U_{\alpha j}(\omega)$. In Eq.~\eqref{eq:Bfinal} the bold symbols denote matrices of dimension $N_{pdep}^2$. 
In order to compute the matrix elements of the correlation self-energy, Eq.~\eqref{eq:SCABCD}, we need to evaluate $\Xi(\omega)$ and $\boldsymbol\Lambda(\omega)$, namely the head and body of the $\bar{\chi}$ operator. These are easily obtained via linear algebra operations from $F_{\alpha\beta}(\omega)$, $U_{\alpha j}(\omega)$, $U_{i\alpha}(\omega)$ and $B_{ij}(\omega)$. \\
By replacing explicit summations over unoccupied states with projection operations, Eq.s~\eqref{eq:finalF}-\eqref{eq:finalB} may be evaluated using linear equation solvers and (owing to the completeness of the energy eigenstates) the calculation of polarizabilities is carried out without the explicit evaluation of the virtual states. In a similar fashion one obtains the auxiliary functions $\xi^\alpha_{n\vc{k}\sigma}(\vc{r})$ in Eq.~\eqref{eq:commu} and the PDEP basis set as described in Sec.~\ref{sec:pdepmethod}. 
We note that other approaches were developed in the literature\cite{Bruneval_2008_extrapolaronapprox,reining_2010_collapsing,Kang_2010_cohsexplus,Samsonidze_2011_sapo} to improve the efficiency of $G_0W_0$ calculations
by avoiding the calculation of virtual states, or by limiting the number of virtual states to be computed. However such approaches did not make use of the spectral decomposition of the irreducible polarizability to obtain the reducible polarizability, but instead inverted explicity large matrices. Specifically in Reining \textit{et al.}\cite{reining_1997_DFPTforchi0} the Sternheimer equation was used to obtain the irreducible polarizability without virtual states and then a plasmon pole model was adopted to compute the dielectric response as a function of frequency. In Giustino \textit{et al.}\cite{Giustino_2013_SternGW} the Sternheimer equation was used as well to obtain the irreducible polarizability without computing virtual states; the polarizability matrix was then inverted numerically and either a plasmon pole model or a a Pad\'e expansion were used to treat the frequency dependence.
In our approach we avoided large matrix inversion by using the PDEP basis set to express all polarizability matrices. Finally, we note that an additional advantage of our approach is that Eq.s~\eqref{eq:finalF}-\eqref{eq:finalB} may be computed using a deflated Lanczos algorithm for multiple values of the frequency, as discussed in Sec.~\ref{sec:lglw}. A Lanczos algorithm was also used by  Soininen \textit{et al.}\cite{soininen_2003_GWLanczos} to iteratively include local field effects in RPA Hamiltonians and avoid explicit inversion of large matrices. However the authors of Ref.~[\onlinecite{soininen_2003_GWLanczos}] computed explicitly virtual states.
%


\subsection{Projective dielectric eigenpotential (PDEP) basis set}
\label{sec:pdepmethod}
\begin{figure}
\includegraphics[width=0.45\textwidth]{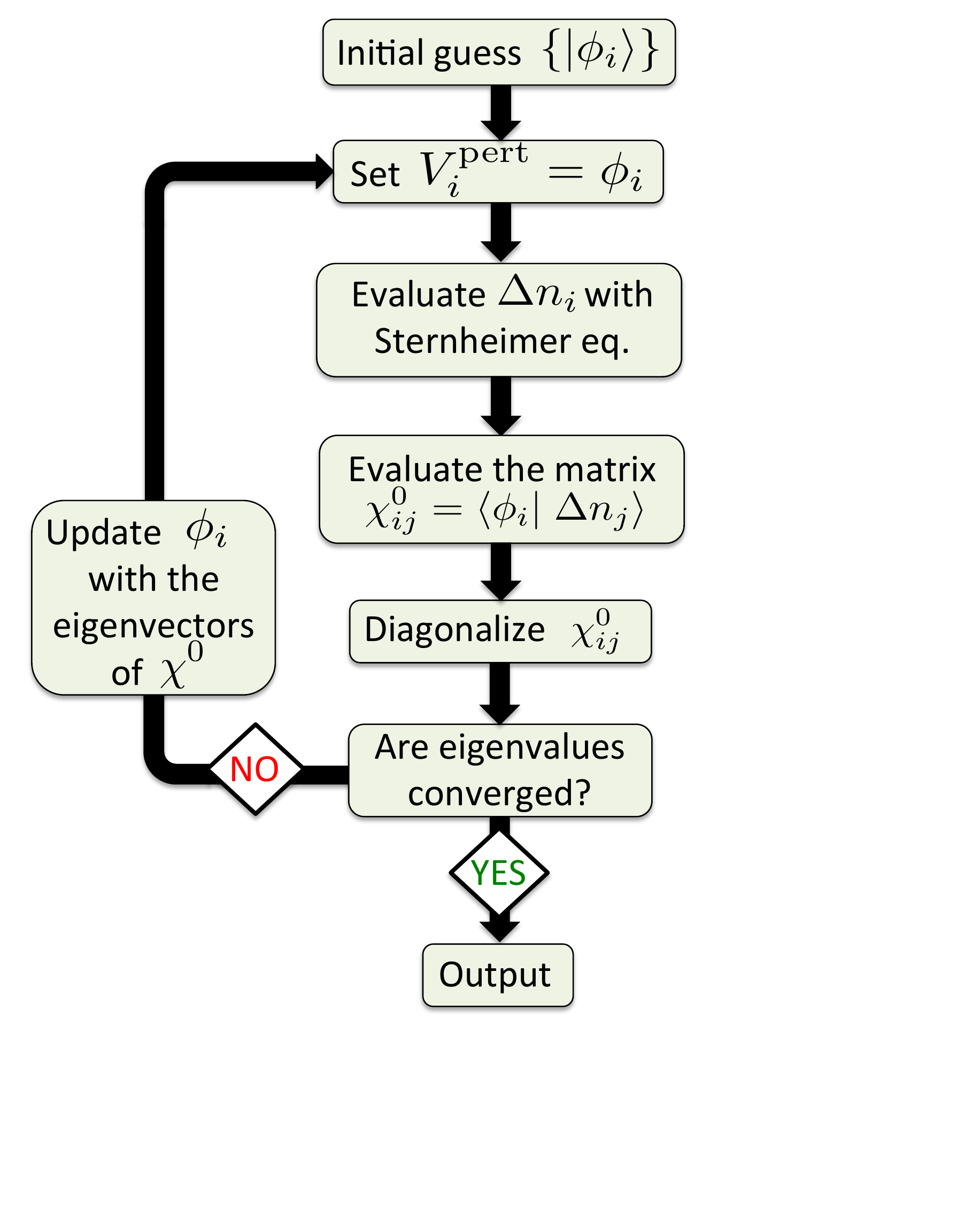}
\includegraphics[width=0.45\textwidth]{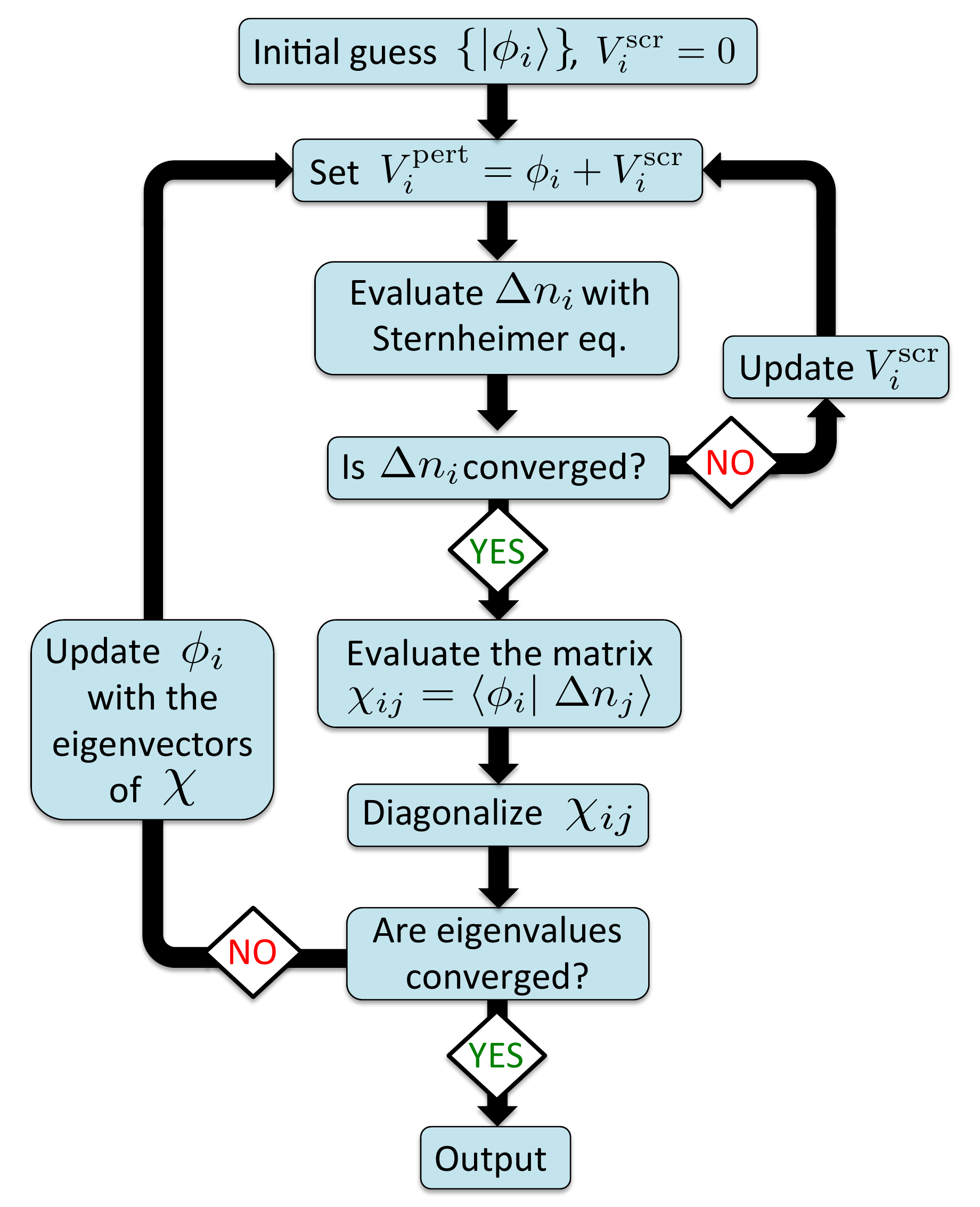}
  \caption{(Color online) Diagrams of the iterative diagonalization of the irreducible ($\chi^0$, left panel) and reducible ($\chi$, right panel) polarizability adopted in this work. In both cases the initial set of vectors $\left\{\ket{\phi_i}\right\}$ are assigned with random components. At each iteration the polarizability matrix is computed by evaluating the density response to the $i$-th perturbation using Eq.\eqref{eq:stern}-\eqref{eq:chiisamatrixinpdep}. The two diagrams differ only by the self-consistent inclusion of $\hat{V}_i^{\text{scr}}$ in the solution of the Sternheimer equation.}
  \label{fig:pdepscheme}
\end{figure}

\begin{figure}
\includegraphics[width=0.45\textwidth]{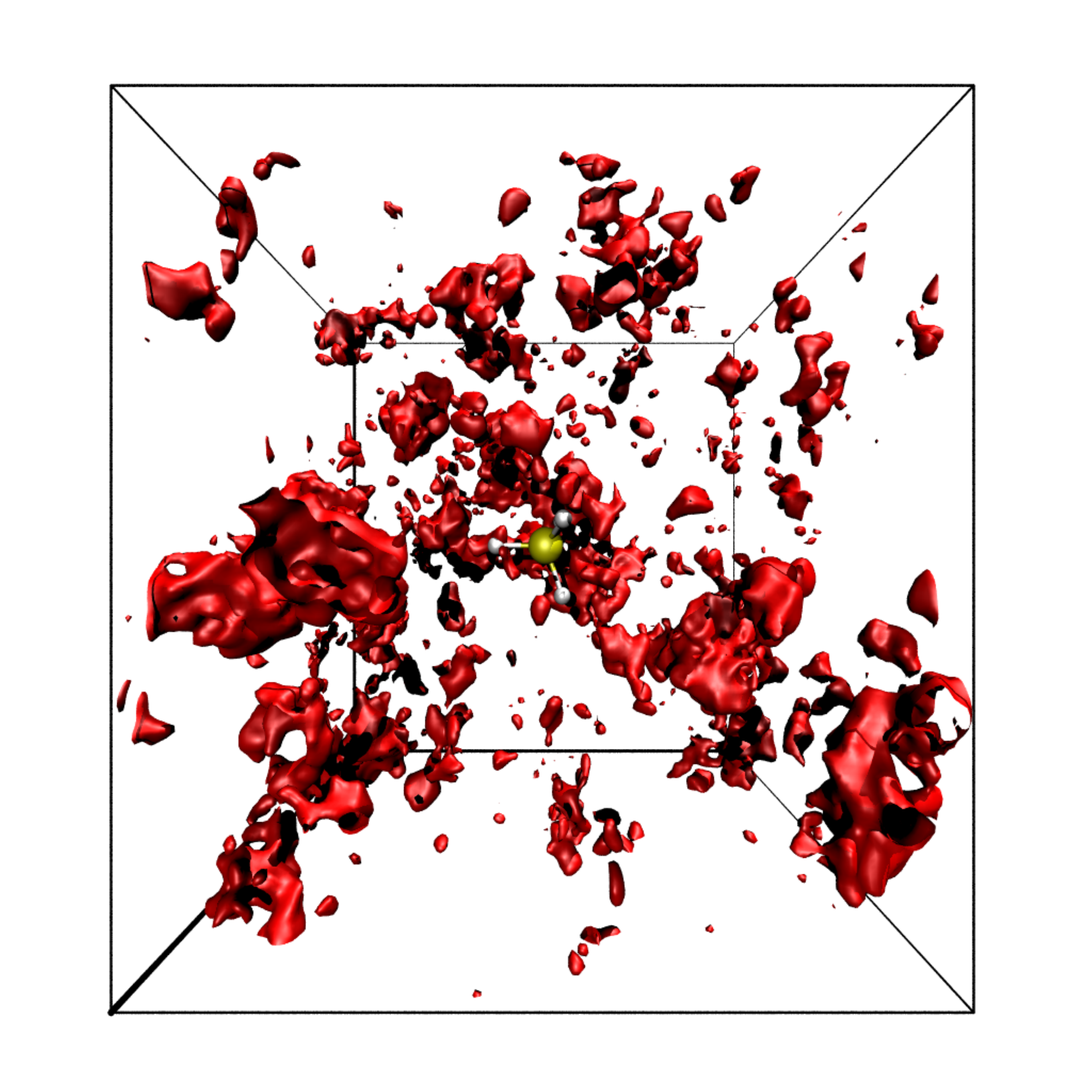}
\includegraphics[width=0.45\textwidth]{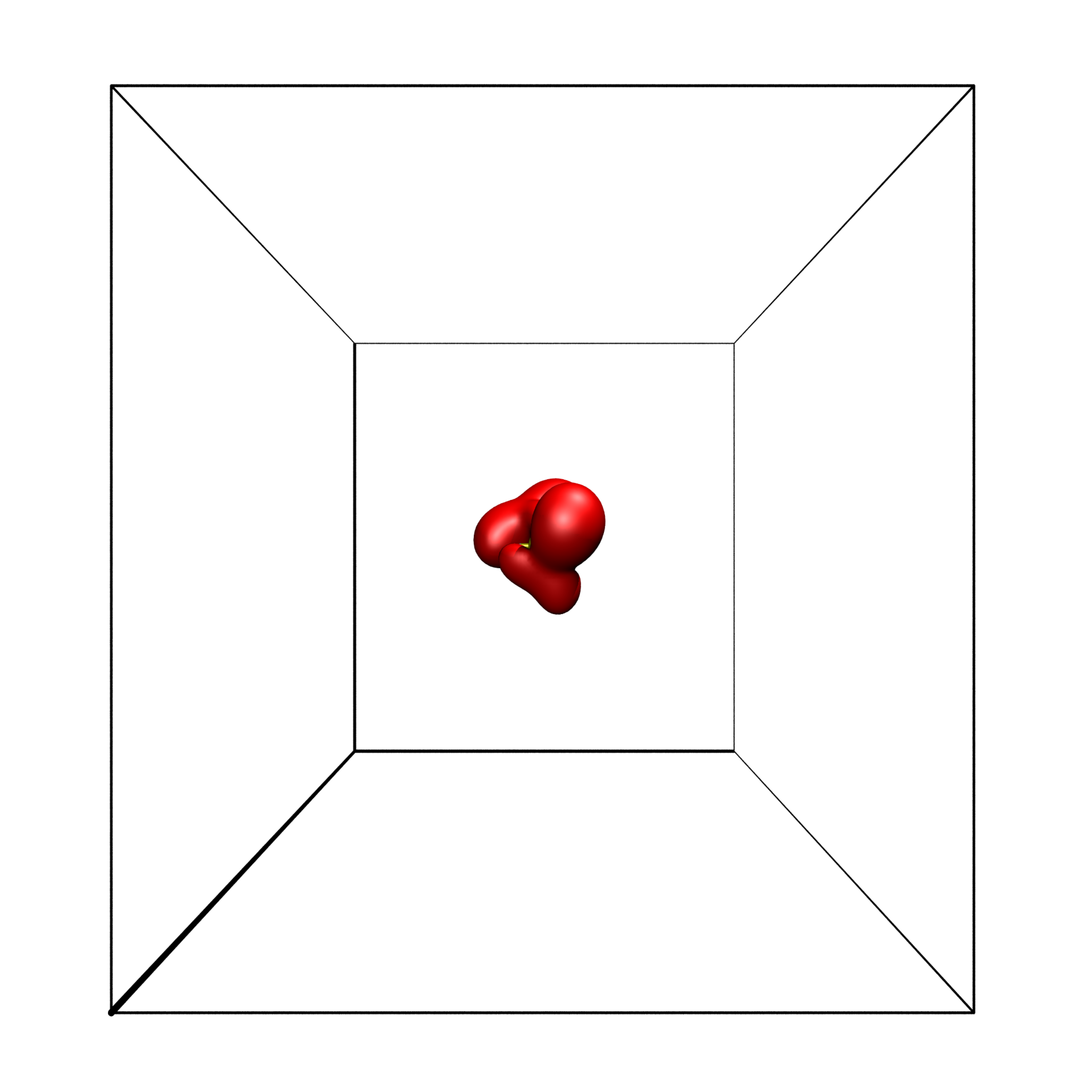}
  \caption{(Color online) Isosurfaces of the square modulus of the most screened eigenvector of the polarizability matrix of the silane molecule ($|\phi_i(\vc{r})|^2$ in Fig.~\ref{fig:pdepscheme}, left panel). The iterative diagonalization was started from vectors with random components (left panel) and converges rapidly (3-4 iterations) to the potential shown in the right panel).} 
  \label{fgr:pdepiter}
\end{figure}

\begin{figure}
\includegraphics[width=0.6\textwidth]{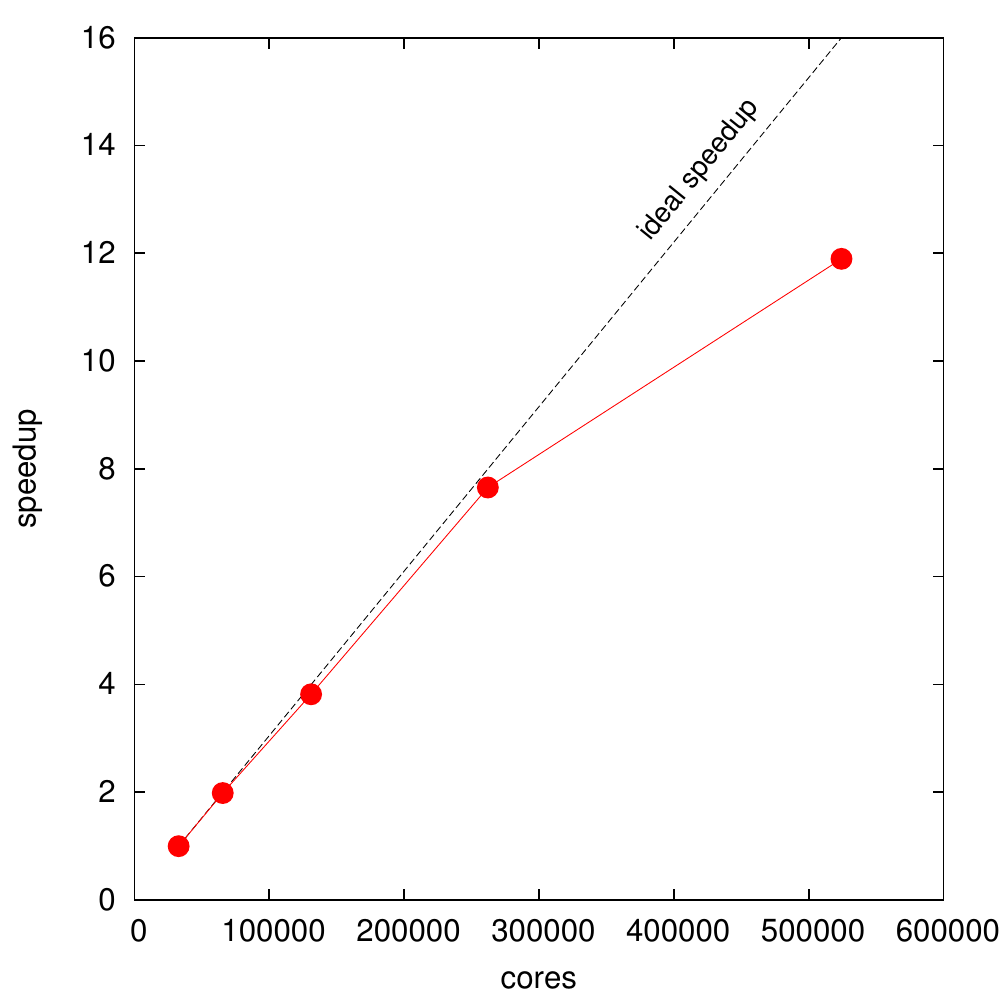}
  \caption{(Color online) Scalability of the PDEP iterative diagonalization (see Fig.~\ref{fig:pdepscheme}) of the static dielectric matrix of the \ce{COOH-Si}/\ce{H_2O} solid/liquid interface discussed in Sec.~\ref{sec:interfaces}. The unit cell includes 492 atoms and 1560 valence electrons.}
  \label{fig:pdep_scalability}
\end{figure}

\begin{figure}
\includegraphics[width=0.95\textwidth]{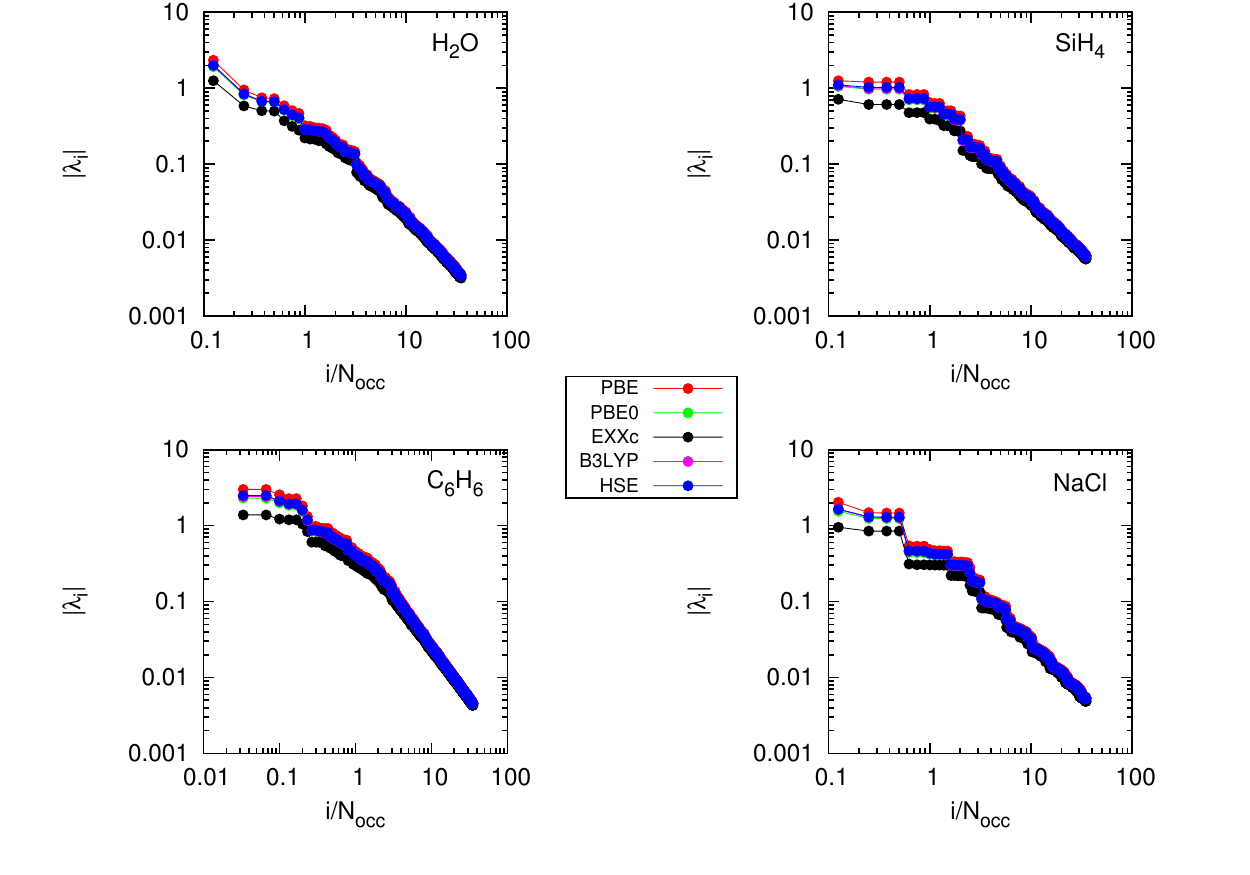}
  \caption{(Color online) Eigenvalues ($\lambda_i$) of the polarizability of \ce{H_2O}, \ce{SiH_4}, \ce{C_6H_6} and \ce{NaCl} molecules, as obtained using the iterative diagonalization described in the left panel of Fig.~\ref{fig:pdepscheme} (see text), and adopting five different exchange-correlation potentials for the KS Hamiltonian (see Table~\ref{tbl:gksrecap}). $N_{occ}$ is the number of valence bands\cite{Wilson_2009_PDEP}.}
  \label{fgr:pdep_hybrid}
\end{figure}

We now describe in detail how to obtain the PDEP basis set $\left\{\ket{\phi_i}\; : \; i=1,N_{pdep}\right\}$ introduced in Eq.~\eqref{eq:eigendecompzerofreq}; each function $\phi_i$ is computed by the iterative diagonalization procedure, summarized in Fig.~\ref{fig:pdepscheme}, the procedure is initiated by building an orthonormal set of $N_{pdep}$ basis vectors, e.g. with random components. Then $N_{pdep}$ Sternheimer equations are solved in parallel, where the perturbation is given by the $i$-th basis set vector $\phi_i(\vc{r})$. In particular, given a perturbation $\hat{V}_i^{\text{pert}}$, the linear variation $\left| \Delta \psi^i_{n\vc{k}\sigma} \right\rangle$ of the occupied eigenstates of the unperturbed system $\left| \psi_{n\vc{k}\sigma} \right\rangle$ may be evaluated using the Sternheimer equation\cite{sternheimer_1954}:
\begin{equation}
\left(\hat{H}^\sigma_{KS} - \varepsilon_{n\vc{k}\sigma}\right) \hat{P}_{c}^{\vc{k}\sigma} \left| \Delta \psi^i_{n\vc{k}\sigma} \right\rangle = - \hat{P}_c^{\vc{k}\sigma} \hat{V}_i^{\text{pert}} \left| \psi_{n\vc{k}\sigma} \right\rangle \,. \label{eq:stern}
\end{equation}
Eq.~\eqref{eq:stern} may be iteratively solved using e.g. preconditioned conjugate-gradient methods. The linear variation of the density due to the $i$-th perturbation is obtained within density functional perturbation theory\cite{baroni_1987_deltavscf,Baroni_2001_revmodphys} (DFPT) as
\begin{equation}
\Delta n_i (\vc{r}) =\sum_\sigma \sum_{n=1}^{N^\sigma_{occ}} \int\limits_{BZ} \frac{d\vc{k}}{(2\pi)^3}  \left[ \Delta \psi^{i\ast}_{n\vc{k}\sigma} (\vc{r}) \psi_{n\vc{k}\sigma} (\vc{r}) + c.c. \right] \,. \label{eq:deltani}
\end{equation}
The matrix elements of the irreducible polarizability in the space spanned by $\phi_i$ are given by:
\begin{equation}
\bar{\chi}^0_{ij}=4\pi e^2 \langle\tilde{\phi}_i\ket{\Delta n_j} \label{eq:chiisamatrixinpdep}
\end{equation}
where $\ket{\Delta n_j}$ is computed using Eq.s~\eqref{eq:stern}-\eqref{eq:deltani} and assuming that $V^{\text{pert}}_i(\vc{G})=\tilde{\phi}_i(\vc{G})$. The matrix $\bar{\chi}^0_{ij}$ is then diagonalized to obtain new $N_{pdep}$ basis vectors $\phi_i$, and the procedure is iterated using e.g. a Davidson algorithm\cite{Davidson_1975_algo} (See Fig~\ref{fgr:pdepiter}). We note that at each iteration, all Sternheimer problems are independent from each other, thus offering the opportunity to carry out embarrassingly parallel calculations. A description of the parallel operations and data layout will be given elsewhere~\cite{Govoni_2014_parallelmethod}. As a result, the algorithm shows a good scalability up to 524288 cores (see Fig.~\ref{fig:pdep_scalability}).\\
As an example we show in Fig.~\ref{fgr:pdep_hybrid} the eigenvalues of the $\bar{\chi}^0_{ij}$ matrix obtained with the PDEP algorithm for the water, silane, benzene and sodium chloride molecules, using KS Hamiltonians with different exchange-correlation functionals. The choice of the functional only affects the most screened eigenpotentials, whereas the eigenvalues $\lambda_i$ corresponding to the least screened ones rapidly approach\cite{Wilson_2008_PDEP} zero with a decay similar to that predicted by the Lindhard model\cite{Wilson_2009_PDEP}. This indicates that the computation of the least screened eigenpotentials might be avoided and carried out using model functions. \\ 
If instead of $\bar{\chi}^0$, one wishes to diagonalize $\bar{\chi}$, the potential $\hat{V}_i^{\text{scr}}$ arising from the rearrangements of the charge density in response to the applied perturbation needs to be included in the definition of the perturbation $\hat{V}_i^{\text{pert}} $ of Eq.~\eqref{eq:stern}\cite{ehrenreich_book,hanke,hybertsen_dm}. In a generalized KS scheme the $\hat{V}_i^{\text{scr}}$ is given by:
\begin{equation}
\hat{V}^\text{scr}_i \left|\psi_{n\vc{k}\sigma} \right\rangle = \left[ \Delta \hat{V}^i_H + \left(1-\alpha \right) \Delta \hat{V}_x^i + \Delta \hat{V}_c^i + \alpha \Delta \hat{V}_{EXX}^i \right ]    \left|\psi_{n\vc{k}\sigma} \right\rangle \label{eq:LF}
\end{equation}
where $\alpha$ is the fraction of exact exchange (EXX) that is admixed to the semilocal exchange potential. The linear variation of the Hartree potential is 
\begin{equation}
\Delta \hat{V}^i_H \left|\psi_{n\vc{k}\sigma} \right\rangle = \int d\vc{r^\prime} \Delta n_i (\vc{r^\prime}) \frac{e^2}{|\vc{r}-\vc{r^\prime}|} \psi_{n\vc{k}\sigma} (\vc{r}) \label{eq:LFHartree}
\end{equation}
and those of the exchange and correlation potentials are given by the functional derivatives:
\begin{equation}
\Delta \hat{V}_{x/c}^i\left|\psi_{n\vc{k}\sigma} \right\rangle = \left. \frac{dV_{x/c}}{dn} \right|_{n(\vc{r})} \Delta n_i (\vc{r}) \psi_{n\vc{k}\sigma} (\vc{r}) \,.
\end{equation}
The linear variation of the exact exchange potential (EXX) is expressed in terms of variations of the single particle orbitals
\begin{equation}
\Delta \hat{V}_{EXX}^i \left|\psi_{n\vc{k}\sigma} \right\rangle = - \sum_{m=1}^{N^{\sigma}_{occ}} \int\limits_{BZ} \frac{d\vc{k^\prime}}{(2\pi)^3}   \int d\vc{r^\prime} \left[ 
\Delta \psi^{i\ast}_{m\vc{k^\prime}\sigma} (\vc{r^\prime}) \psi_{m\vc{k^\prime}\sigma} (\vc{r}) + 
\psi^\ast_{m\vc{k^\prime}\sigma} (\vc{r^\prime}) \Delta \psi^{i}_{m\vc{k^\prime}\sigma} (\vc{r}) 
\right] \frac{e^2}{|\vc{r}-\vc{r^\prime}|} \psi_{n\vc{k}\sigma} (\vc{r^\prime}) \,.
\end{equation}
We note that calculations including $\hat{V}_i^{\text{scr}}$ require a double self-consistent procedure (see Fig.~\ref{fig:pdepscheme}); hence it is computationally more efficient to iteratively diagonalize $\bar{\chi}^0$ first and then obtain the reducible polarizabilty $\bar{\chi}$ by linear algebra operations. \bibnote{Because of the RPA, Eq.~\eqref{eq:linearalgebrachichi0} formally corresponds to the solution of $\bar{\chi}$ when only the Hartree contribution of Eq.~\eqref{eq:LFHartree} is considered in Eq.~\eqref{eq:LF}}. We recall that both $\bar{\chi}$ and $\bar{\chi}^0$ are Hermitian operators\cite{Car_1981_dbs} and because of Eq.~\eqref{eq:linearalgebrachichi0} they have the same eigenvectors. 

%
%
%
%
%


 \subsection{The evaluation of $G$ and $W$ without empty electronic states using a Lanczos algorithm}
\label{sec:lglw}
The calculation of the correlation self-energy, Eq.s~\eqref{eq:SCA}-\eqref{eq:SCD}, and of the screening, Eq.s~\eqref{eq:finalF}-\eqref{eq:finalB}, requires the computation of the matrix elements of $\hat{O}^{\sigma}_{KS}(\omega)$, defined in Eq.~\eqref{eq:Odef}, for multiple values of $\omega$. For each frequency $\omega$, given two generic vectors $\ket{L}$ and $\ket{R}$, we define
\begin{equation}
M^{\vc{k}\sigma}_{v;LR}(\omega) = \bra{L} \hat{P}_v^{\vc{k}\sigma} \hat{O}^{\sigma}_{KS}\left( \omega\right) \hat{P}_v^{\vc{k}\sigma} \ket{R}  \label{eq:MV}
\end{equation}
and 
\begin{equation}
M^{\vc{k}\sigma}_{c;LR}(\omega) = \bra{L} \hat{P}_c^{\vc{k}\sigma} \hat{O}^{\sigma}_{KS}\left( \omega \right) \hat{P}_c^{\vc{k}\sigma} \ket{R}.   \label{eq:MC}
\end{equation}
Eq.~\eqref{eq:MV} can be easily written in terms of the eigenstates $\psi_{n\vc{k}\sigma}$ and eigenvalues $\varepsilon_{n\vc{k}\sigma}$ of the KS Hamiltonian:
\begin{equation}
M^{\vc{k}\sigma}_{v;LR}(\omega) = - \sum_{i=1}^{N^\sigma_{occ}} \frac{ \braket{L}{\psi_{n\vc{k}\sigma}} \braket{\psi_{n\vc{k}\sigma}}{R} }{\varepsilon_{n\vc{k}\sigma} -\omega} \label{eq:MVfinal}
\end{equation}
Eq.~\eqref{eq:MC} may be cast as well in terms of the occupied states and energies, by using the relation $P_c^{\vc{k}\sigma}=1-P_v^{\vc{k}\sigma}$ and writing $\tilde{H}^{\sigma}_{KS}=\hat{P}_c^{\vc{k}\sigma} \hat{H}^{\sigma}_{KS} \hat{P}_c^{\vc{k}\sigma}$ (called deflated Hamiltonian) 
\begin{equation}
M^{\vc{k}\sigma}_{c;LR}(\omega) = - \langle L| \left( \tilde{H}^{\sigma}_{KS} -\omega \right)^{-1} | R\rangle \label{eq:MCdef}
\end{equation}
The Lanczos alorithm\cite{cini_book} is used to obtain a set of $N_{lanczos}$ orthonormal vectors $Q=\left\{ \ket{q_i} \; :\; i=1,N_{lanczos}\right\}$ that are used to recast the deflated Hamiltonian in Eq.~\eqref{eq:MCdef} into a tri-diagonal form:
\begin{equation}
T=Q^\dagger \tilde{H}^{\sigma}_{KS} Q = 
\begin{pmatrix}
\alpha_1 & \beta_2 & & & \\ 
\beta_2 & \alpha_2 & \beta_3 & & \\
& \beta_3 & \ddots & \ddots & \\
& & \ddots & \ddots & \beta_{n} \\
& & & \beta_{n} & \alpha_n 
\end{pmatrix}
\end{equation}
where 
\begin{equation}
\alpha_n = \bra{q_n}  \tilde{H}^{\sigma}_{KS} \ket{q_n}\,,
\end{equation}
\begin{equation}
\beta_n = \parallel  (\tilde{H}^{\sigma}_{KS} - \alpha_n) \ket{q_n} - \beta_n \ket{q_{n-1}} \parallel \,.
\end{equation}
The calculation of the sequence of vectors $\ket{q_n}$, called Lanczos chain, is started by imposing $\ket{q_1}=\ket{R}$; iterations are performed\bibnote{For the systems presented in this work, $N_{Lanczos}=40$ yielded converged results.} by  enforcing orthogonality through the recursive relation:
\begin{equation}
\ket{q_{n+1}} = \frac{1}{\beta_{n+1}} \left[ (\tilde{H}^{\sigma}_{KS} - \alpha_n) \ket{q_n} - \beta_n \ket{q_{n-1}}  \right]\,.
\end{equation}
The tridiagonal matrix $T$ can be diagonalized using an orthogonal transformation $U$, so that $D=U^t T U$. Using $d_n$ to indicate the $n$-th element of the diagonal matrix $D$, we obtain:
\begin{equation}
M^{\vc{k}\sigma}_{c;LR}(\omega) = - \sum_{\substack{n_1=1,\\n_2=1,\\n_3=1}}^{N_{lancsoz}} \braket{L}{q_{n_1}} U_{n_1n_2} \frac{1}{d_{n_2}-\omega} U_{n_3n_2} \braket{q_{n_3}}{R} \,.
\end{equation}
Because of the orthogonality of the elements belonging to a Lanczos chain, we have $\braket{q_{n_3}}{R} = \delta_{n_31}$, yielding
\begin{equation}
M^{\vc{k}\sigma}_{c;LR}(\omega) = - \sum_{\substack{n_1=1,\\n_2=1}}^{N_{lancsoz}} \braket{L}{q_{n_1}} U_{n_1n_2} \frac{1}{d_{n_2}-\omega} U_{1n_2} \,. \label{eq:MCfinal}
\end{equation}
Eq.~\eqref{eq:MCfinal} is written in a form similar to Eq.~\eqref{eq:MVfinal}, where the coefficients of the expansion are independent of the value of $\omega$ and therefore it is not necessary to recompute them for multiple frequencies. However, the coefficients of the expansion in Eq.~\eqref{eq:MCfinal} depend by construction on the vector $\ket{R}$ that is used to start the Lanczos chain. Therefore to evaluate $M^{\vc{k}\sigma}_{c;LR}(\omega)$ one needs to solve as many Lanczos problems as the number of vectors $\ket{R}$, while the eigendecoposition used for $M^{\vc{k}\sigma}_{v;LR}(\omega)$ in Eq.~\eqref{eq:MVfinal} is unique. Because Lanczos chains are independent of each other, the iterations can be performed in an embarrassingly parallel fashion, similarly to the procedure discussed in Sec.~\ref{sec:pdepmethod} for the computation of the PDEP basis set. \\
In our calculations we used Eq.~\eqref{eq:MVfinal}, with an explicit summation over occupied eigenstates, for the evaluation of the terms in Eq.~\eqref{eq:SCA} and Eq.~\eqref{eq:SCC}, whereas we used the Lanczos expansion of Eq.~\eqref{eq:MCfinal} to evaluate the terms in Eq.s~\eqref{eq:SCB}, \eqref{eq:SCD}, \eqref{eq:finalF}-\eqref{eq:finalB}. 


\subsection{The contour deformation technique}
\label{sec:contourdef}
\begin{figure}
\includegraphics[width=0.6\textwidth]{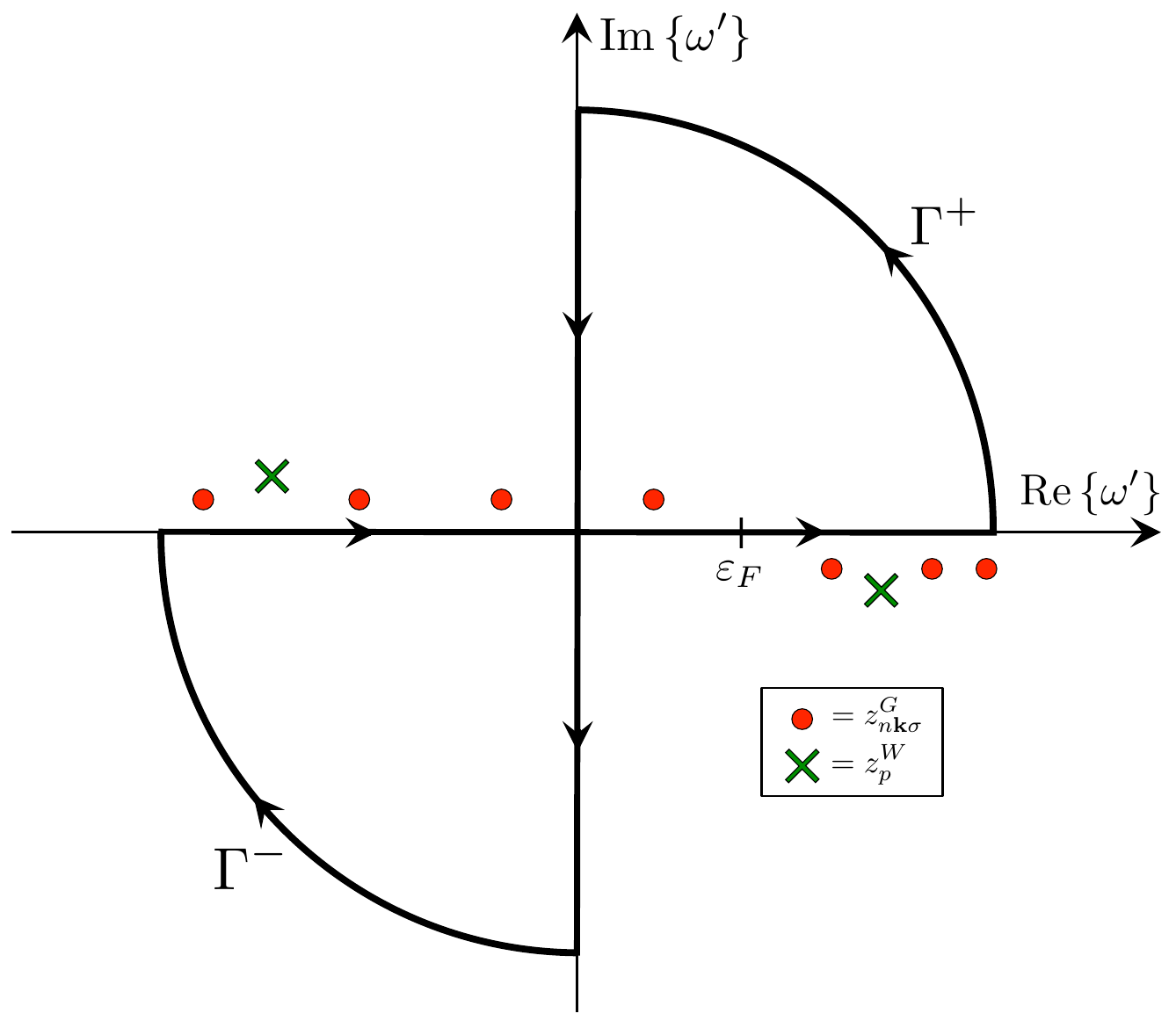}
  \caption{(Color online) Contours used in this work (see text). The integration contours $\Gamma^+$ and $\Gamma^-$ enclose only the poles of the Green's function $z^{G}_{n\vc{k}\sigma}$ (dots) and exclude the poles of the screened Coulomb interaction $z^{W}_{p}$ (crosses).}
  \label{fig:contourSc}
\end{figure}

In Eq.s~\eqref{eq:SCABCD}-\eqref{eq:SCD} the frequency integration may be carried out by using complex analysis, and thus avoiding the integration in the real frequency domain. A closed integration contour on the complex plane is identified for which Cauchy's theorem and Jordan's Lemma apply\cite{dennery_book}. This approach is called contour deformation technique~\cite{godby_qp,giantomassi_2011_interfaces} and establishes a formal identity between the quantities reported in Eq.~\eqref{eq:SCABCD} and an equivalent set of quantities that are numerically more stable to compute. The poles of the Green's function $G_{KS}^\sigma(\vc{r},\vc{r}^\prime ; \omega+\omega^\prime )$ are located at complex frequencies $z^G_{n\vc{k}\sigma}$, satisfying the relation
\begin{equation}
z^G_{n\vc{k}\sigma} = \varepsilon_{n\vc{k}\sigma}-\omega - i \eta \text{sign}( \epsilon_{n\vc{k}\sigma} - \epsilon_F)
\end{equation}
with a numerical residue given by
\begin{equation}
\text{Res}\left\{ G^\sigma_{KS} (\vc{r},\vc{r^\prime}), z^G_{n\vc{k}\sigma}  \right\} =  \psi_{n\vc{k}\sigma}(\vc{r})\psi^\ast_{n\vc{k}\sigma}(\vc{r^\prime}) .
\end{equation}
The poles of $W_p$ correspond to the plasmon energies of the system: $z^{W}_p=\pm(\Omega_p-i\eta)$. 
The matrix elements of the correlation self-energy can be computed by using the integration contour shown in Fig.~\ref{fig:contourSc}, yielding:
 \begin{eqnarray}
\Sigma^\sigma_C(\vc{r},\vc{r^\prime};\omega) &=& i \int\limits_{-i\infty}^{+i\infty} \frac{d\omega^\prime}{2\pi} G^\sigma_{KS}(\vc{r},\vc{r^\prime};\omega+\omega^\prime) W_p(\vc{r},\vc{r^\prime};\omega^\prime) +\\
&&- \sum_{z^{G}_{n\vc{k}\sigma}\in\Gamma^{+}}  \psi_{n\vc{k}\sigma}(\vc{r})\psi^\ast_{n\vc{k}\sigma}(\vc{r^\prime})  W_p(\vc{r},\vc{r^\prime};z^{G}_{n\vc{k}\sigma}) +\\
&&+ \sum_{z^{G}_{n\vc{k}\sigma}\in\Gamma^{-}}  \psi_{n\vc{k}\sigma}(\vc{r})\psi^\ast_{n\vc{k}\sigma}(\vc{r^\prime}) W_p(\vc{r},\vc{r^\prime};z^{G}_{n\vc{k}\sigma})\,.
\end{eqnarray}

In view of the chosen contour, as the frequency $\omega$ is varied, the poles of $W_p$ never fall inside the two closed contours $\Gamma^+$ and $\Gamma^-$, which therefore may only enclose poles of the Green's function. The correlation self-energy is thus obtained as the sum of: i) an integral along the imaginary axis, where both $G^\sigma_{KS}$ and $W_{RPA}$ are smooth functions, and ii) all the numerical residues arising from $G^\sigma_{KS}$, shifted inside the integration contours, depending on the value of $\omega$. 
The matrix element of the correlation self-energy becomes:
\begin{eqnarray}
\bra{\psi_{n\vc{k}\sigma}} \Sigma^\sigma_C(E^{QP}_{n\vc{k}\sigma}) \ket{\psi_{n\vc{k}\sigma}} &=& -\int\limits_{-\infty}^{+\infty} \frac{d\omega^\prime}{2\pi} \bra{\psi_{n\vc{k}\sigma}} G^\sigma_{KS}(\vc{r},\vc{r^\prime};E^{QP}_{n\vc{k}\sigma}+i\omega^\prime) W_p(\vc{r},\vc{r^\prime};i\omega^\prime) \ket{\psi_{n\vc{k}\sigma}} +\\
&&+ \sum_{m} f_{m\vc{k}\sigma}^{n\vc{k}\sigma}    \bra{\psi_{n\vc{k}\sigma}}  \psi_{m\vc{k}\sigma}(\vc{r})  W_p(\vc{r},\vc{r^\prime};\varepsilon_{m\vc{k}\sigma}-E^{QP}_{n\vc{k}\sigma}) \psi^\ast_{m\vc{k}\sigma}(\vc{r^\prime}) \ket{\psi_{n\vc{k}\sigma}} \nonumber \label{eq:contourdeformationfinal}
\end{eqnarray}
where the function $f_{m\vc{k}\sigma}^{n\vc{k}\sigma}$ is 
\begin{equation}
f_{m\vc{k}\sigma}^{n\vc{k}\sigma} = 
\begin{cases}
+1 & \text{if} \;\;\; \varepsilon_F < \varepsilon_{m\vc{k}\sigma} < E^{QP}_{n\vc{k}\sigma} \\
-1 & \text{if} \;\;\; E^{QP}_{n\vc{k}\sigma} < \varepsilon_{m\vc{k}\sigma}< \varepsilon_F \\
0 & \text{otherwise}
\end{cases} \label{eq:fcontour}
\end{equation}
Eq.~\eqref{eq:fcontour} shows that only a finite number of residues needs to be computed\bibnote{
We note that the GW technique reported in this work replaces explicit summations over empty states present in both $G^\sigma_{KS}$ and $W_p$ with projection operations, thus avoiding the calculation of slow converging summations over empty states. 
However, within first order perturbation theory, in order to get the QP corrections to the energy of the KS state $\psi_{n\vc{k}\sigma}$, one needs to compute the mean value of the self-energy operator over that specific state and hence the wave function $\psi_{n\vc{k}\sigma}$ needs to be computed. 
Eq.~\eqref{eq:fcontour} shows that only the virtual states with energy $\varepsilon_{m\vc{k}\sigma}$ between $\varepsilon_F$ and $E^{QP}_{n\vc{k}\sigma}$ are required, which are available at no extra cost after $(\varepsilon,\psi)_{n\vc{k}\sigma}$ have been obtained from the solution of the KS equations.}.
Inserting Eq.~\eqref{eq:Wchihb} for $W_p$ into Eq.~\eqref{eq:contourdeformationfinal}, Eq.s~\eqref{eq:SCABCD}-\eqref{eq:SCD} are solved.
The integration over the immaginary axis is performed numerically by considering a non-uniform grid, finer for small frequencies. With the introduction of the contour deformation technique we avoid the use of plasmon pole models\cite{hybertsen_qp,godby_1989_PPM,engel_1993_PPM,Aryasetiawan_1998_gw,Shaltaf_2008_PPM,Stankovski_2011_PPM} to describe the frequency dependence of the screening and we overcome the limitations of the analytic continuation\cite{Rieger_1999_GW,giustino_2010_stern} reported in Ref.~[\onlinecite{Viet_2012_GW,Anh_2013_GW}].


\section{Verification and validation of results}
\label{sec:validation}

In this section we present several results obtained with the $G_0W_0$ method presented in Sec.~\ref{sec:methodintro}. In particular we computed the vertical ionization potential (VIP) of closed and open-shell molecules and the electronic structure of several crystalline, amorphous and liquid systems. All results were obtained by computing KS eigenvalues and eigenvectors with the \texttt{QuantumEspresso} package\cite{giannozzi} and the $G_0W_0$ quasiparticle energies with the \texttt{West} code, which features a parallel implementation of the method of Sec.~\ref{sec:methodintro}.

\subsection{Vertical ionization potentials of molecules}
\label{sec:VIPmol}

\begin{figure}
\includegraphics[width=1.0\textwidth]{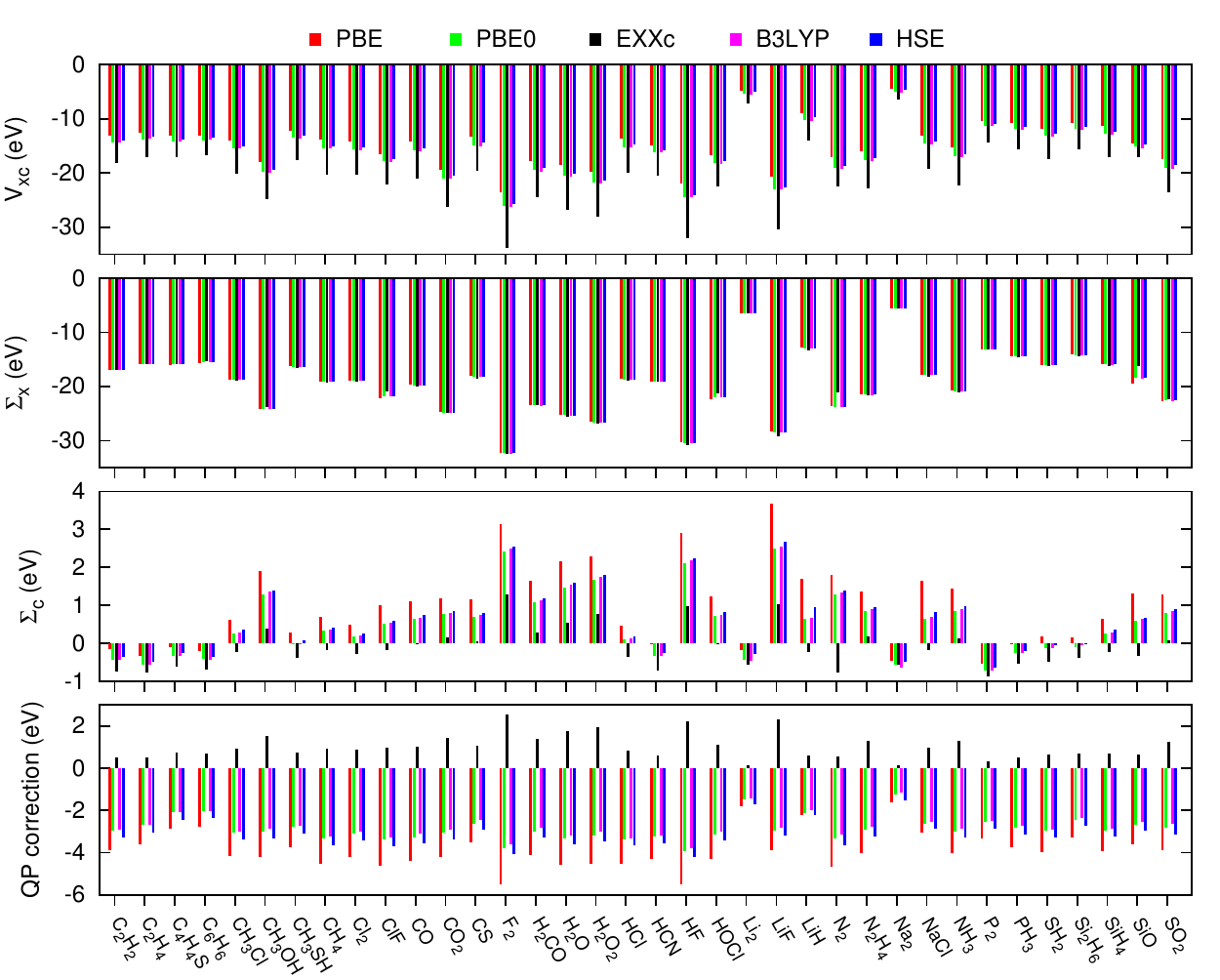}
  \caption{(Color online) The matrix elements of $V_{xc}$, $\Sigma_x$ and $\Sigma_c(E^{QP}_{n\vc{k}\sigma})$ evaluated on the HOMO eigenstate, for different choices of the exchange and correlation potential (see Tab.~\ref{tbl:gksrecap}). The bottom panel reports the QP correction, i.e. the difference $E^{QP}_{n\vc{k}\sigma}-\varepsilon_{n\vc{k}\sigma}$ (see Eq.s~\eqref{eq:QPenergies},~\eqref{eq:fock}-\eqref{eq:sigmac}).}
  \label{fgr:molsummaryBD}
\end{figure}

\begin{figure}
\includegraphics[width=0.65\textwidth]{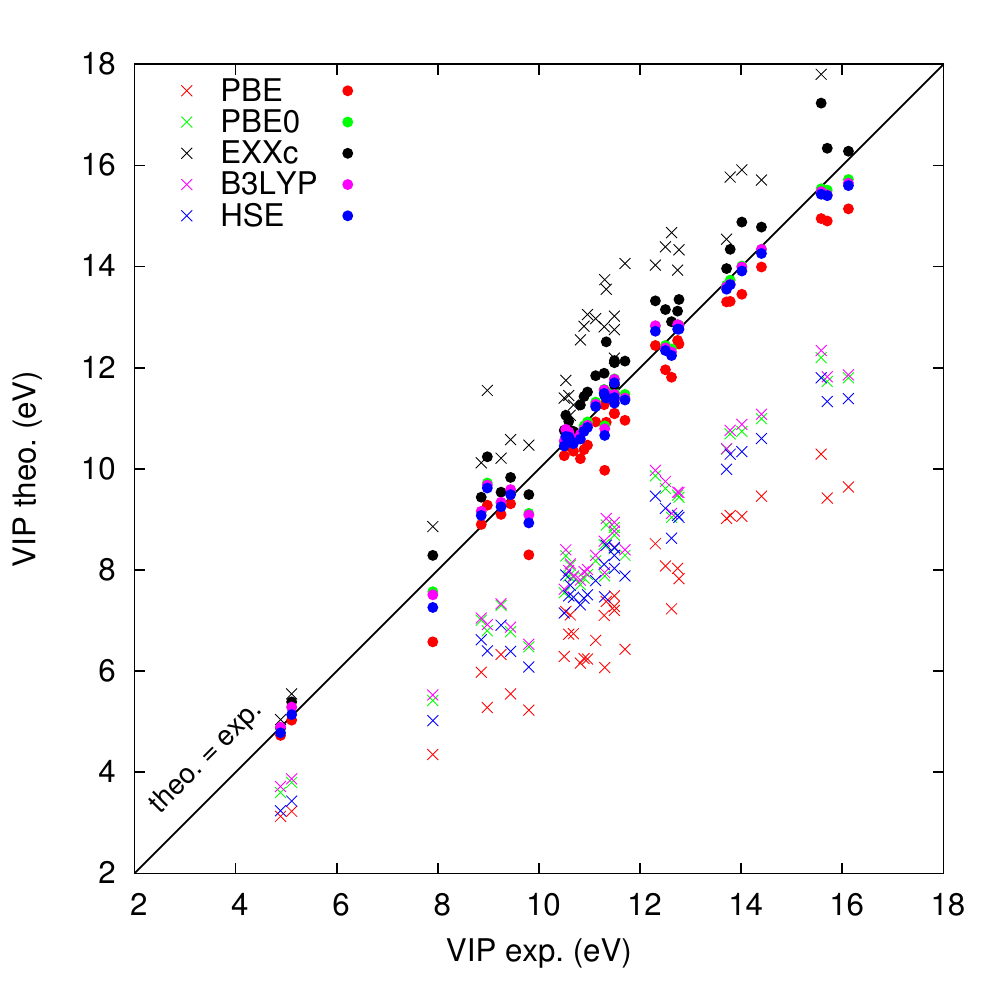}
  \caption{(Color online) Comparison between calculated and experimental vertical ionization potential (VIP) for the set of 36 closed-shell molecules listed in Tab.~\ref{tbl:vipGW}. Dots (crosses) refer to VIPs obtained at the $G_0W_0$ (DFT) level of theory.}
  \label{fgr:molsummaryQP}
\end{figure}

\begin{figure}
\includegraphics[width=1.0\textwidth]{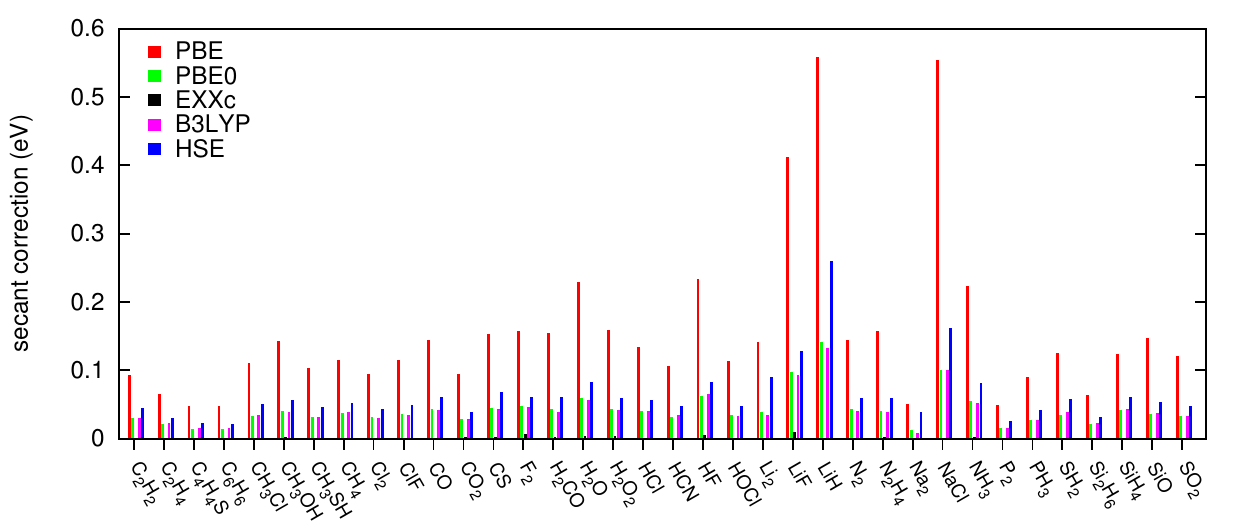}
  \caption{(Color online) Difference between the solution of Eq.~\eqref{eq:QPenergies} using a secant algorithm and employing the first order Taylor expansion of Eq.~\eqref{eq:QPconZ}.}
  \label{fgr:secant}
\end{figure}

We considered a subset of the G2/97 test set\cite{curtiss_1998_g297} composed of 36 closed shell molecules, listed in Table~\ref{tbl:vipGW}. Subsets of the G2/97 set were recently used to benchmark $G_0W_0$ calculations with both localized\cite{Rostgaard_2010_gwmol,Caruso_2010_gwmol,Blase_2011_gw,Ren_2012_GWmol,Bruneval_2013_GW} and plane wave\cite{Sharifzadeh_2012_vip,Anh_2013_GW} basis sets. 
Molecular geometries were taken from the NIST computational chemistry database\cite{nist},  and no  additional structural relaxations were performed. In our calculations, each molecule was placed in a periodically repeated cell of edge $30\,\text{bohr}$; the interaction between ionic cores and valence electrons was described  by a PBE norm conserving pseudopotential; we used a plane wave basis set with a kinetic energy cutoff of $85\,\text{Ry}$ (chosen so as to be appropriate for the hardest pseudopotential, i.e. those of oxygen and fluorine). At the DFT-KS level of theory we approximated the VIP by the absolute value of the highest occupied KS eigenvalue (HOMO)\bibnote{In order to prevent the occurrence of spurious size effects in the computation of VIPs, we referred the KS eigenvalues to the vacuum level. The vacuum level was computed as the spherical average of the electrostatic potential over a sphere of diameter equal to the edge of the cubic simulation cell and centered on the ionic pseudo-charge position. Because in our $G_0W_0$ calculations we did not update the wavefunctions and thus the charge density, the vacuum level was not recomputed when QP corrections were added to KS eigenvalues.} and we considered five different exchange and correlation functionals: PBE, PBE0, EXXc, B3LYP and HSE, whose expressions are summarized in Table~\ref{tbl:gksrecap}. The computed DFT-KS VIP are reported in Table~\ref{tbl:vipGW} within parentheses and in Fig.~\ref{fgr:molsummaryQP} as crosses.  
As expected, hybrid functionals yielded  a better agreement with experiments than PBE: the mean absolute relative errors (MARE) are $13.00\%$, $24.70\%$, $25.51\%$ and $29.22\%$ for EXXc, B3LYP, PBE0 and HSE respectively, whereas the MARE of PBE is substantially higher, $37.98\%$.
\\
Corrections to the DFT eigenvalues were computed within the $G_0W_0$ approximation using the 5 different starting points obtained with the various functionals. The PDEP basis set of each system was generated including a number of eigenpotentials $N_{pdep}$ proportional to the number of valence electrons, for instance $N_{pdep}=1050$ for the largest molecule considered here, i.e. \ce{C_6H_6}. 
The $G_0W_0$ corrected VIPs are reported in Table~\ref{tbl:vipGW} and in Fig.~\ref{fgr:molsummaryQP} as dots; we obtained values in much better agreement with experiments, with MARE of $1.78\%$, $1.96\%$, $2.03\%$, $3.96\%$ and $4.49\%$ for PBE0, B3LYP, HSE, PBE and EXXc starting points, respectively. We note that the QP corrections  to  HOMO DFT eigenvalues have different signs, depending on the starting point: the corrections lead to a  decrease of the VIPs obtained with EXXc  but to an increase of those computed with the other functionals.  In Fig.~\ref{fgr:molsummaryBD}  we analyzed separately the matrix elements of $V_{xc}$, $\Sigma_X$ and $\Sigma_C$  (see Eq.~\eqref{eq:QPenergies}); the latter is the most affected by the choice of the exchange correlation functional at the DFT level.
The matrix elements of $\Sigma_X$ (panel $a$) are instead weakly affected by the choice of the  starting point.\\
In many papers appeared in the literature\cite{Blase_2011_gw,Bruneval_2013_GW}, Eq.~\eqref{eq:QPenergies} is solved using  a linear approximation\cite{Aryasetiawan_1998_gw}:
\begin{equation}
E^{QP}_{n\vc{k}\sigma} \approx  
\epsilon_{n\vc{k}\sigma} +  Z_{n\vc{k}\sigma} \left[ 
\bra{\psi_{n\vc{k}\sigma}} \hat{\Sigma}^\sigma (\varepsilon_{n\vc{k}\sigma}) \ket{\psi_{n\vc{k}\sigma}} - \bra{ \psi_{n\vc{k}\sigma} } \hat{V}_{xc}^\sigma  \ket{\psi_{n\vc{k}\sigma}} \right ]
\label{eq:QPconZ}
\end{equation}
where $Z^{-1}_{n\vc{k}\sigma}=1-\left. \frac{\partial}{\partial\omega} \bra{\psi_{n\vc{k}\sigma}} \hat{\Sigma}^\sigma (\omega) \ket{\psi_{n\vc{k}\sigma}} \right|_{\omega=\varepsilon_{n\vc{k}\sigma}} $. 
Here we employed instead a secant method to find the roots of Eq.~\eqref{eq:QPenergies}, where Eq.~\eqref{eq:QPconZ} was used to determine the starting point of the iterative procedure. The difference between VIPs obtained with Eq.~\eqref{eq:QPconZ} and using the secant method varies within  $0-0.5\,\text{eV}$, see Fig.~\ref{fgr:secant}. \\
%
We also considered five open shell molecules, including the \ce{O_2} molecule in its triplet ground state. The VIPs computed at the DFT level using LDA or the PBE exchange-correlation functional\bibnote{For open shell molecules, the VIPs obtained with the LDA (PBE) exchange correlation functional have been obtained using LDA (PBE) norm-conserving pseudopotentials. Simulations were performed considering collinear spins, a cell of edge $30$ bohr and a kinetic energy cutoff of $85$ Ry.}, and by computing the QP corrections with $G_0W_0$ are summarized in Table~\ref{tbl:openshellmol}. The $G_0W_0$ results are in satisfactory agreement with the experimental data\cite{nist} and, for the systems considered here, LDA seems to provide a better starting point for $G_0W_0$ than PBE.

\begin{figure}
\includegraphics[width=0.78\textwidth]{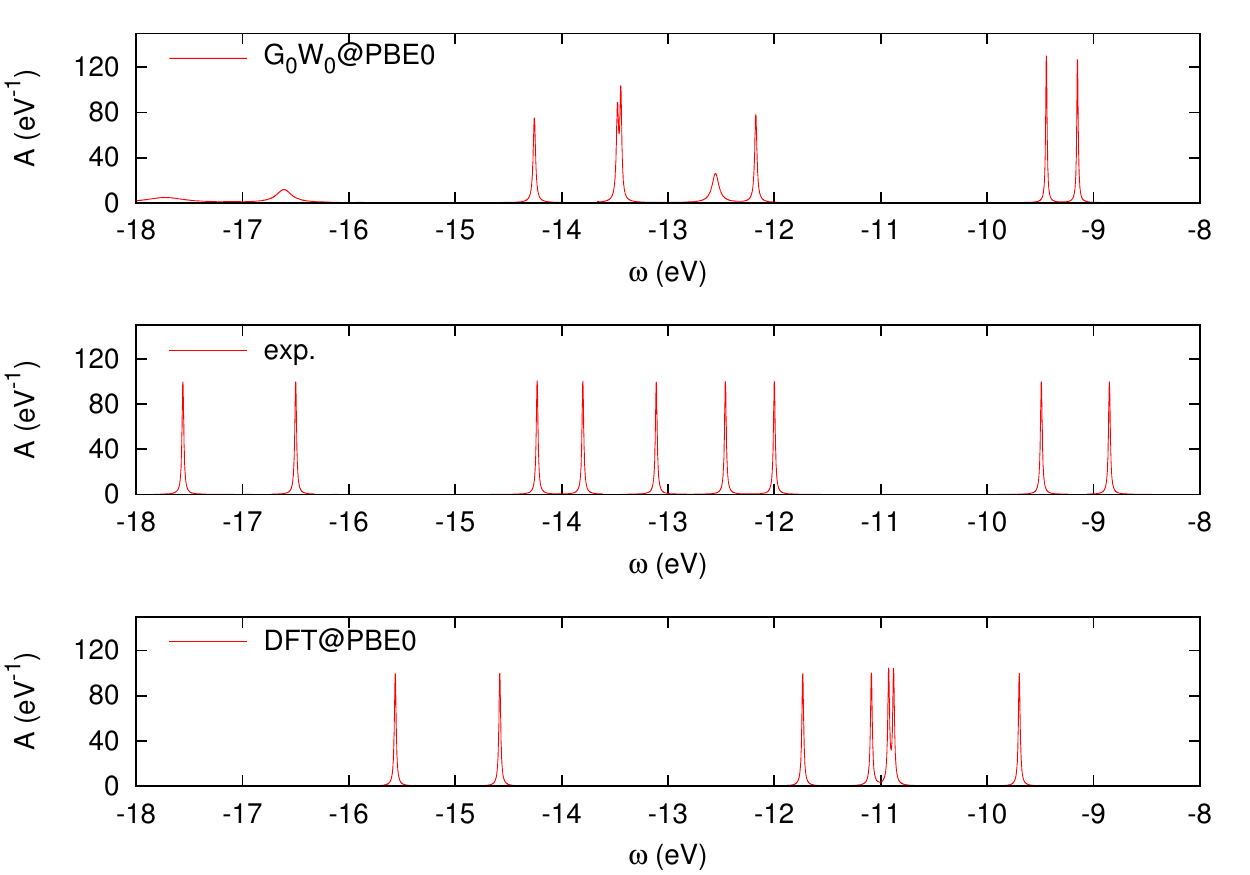}
  \caption{(Color online) The spectral function $A(\omega)$ (see text) for the thiophene molecule (\ce{C4H4S}). The peaks reported in the middle panel are located at the measured ionization potentials\cite{thiophene_ips}. The top (bottom) panel shows the spectral functions obtained at the $G_0W_0$ (DFT) level of theory, using the PBE0 exchange and correlation potential. The width of the peaks is set equal to $0.01\,\text{eV}$, except for the top panel where electronic lifetimes are computed as imaginary part of the self-energies.}
  \label{fgr:lifetimesC4H4S}
\end{figure}

We conclude this section dedicated to the validation of the \texttt{West} code for molecular systems by showing that $G_0W_0$ corrections may also improve higher order VIPs, i.e. vertical ionization energies obtained by removing electrons from single particle states deeper in energy than the HOMO. As an example we chose the thiophene (\ce{C4H4S}) molecule whose spectral function $A(\omega)=\left|\frac{1}{\pi}\text{Tr}\left\{\text{Im}G_0(\omega)\right\}\right|$ was computed within $G_0W_0$, starting from DFT energies obtained using the PBE0 functional (see Fig.~\ref{fgr:lifetimesC4H4S}).
We found that $G_0W_0$ gives a much improved description of higher order VIPs with respect to KS-DFT. While for the experimental and  the PBE0 spectral functions we used an artificial smearing parameter of $\eta = 0.01\,\text{eV}$ to simulate finite lifetimes, in the case of the $G_0W_0$ spectral function the electronic lifetimes were computed from first principles, as the imaginary part of the electron self-energy. Our results are in good agreement with those reported  by F.~Caruso \textit{et al.}\cite{Caruso_2013_scfGW} using localized basis sets.

\begin{table}
\footnotesize
  \caption{Exchange and correlation potentials used in this work (see text). For HSE, the screening parameter $\mu=0.106\,\text{bohr}^{-1}$ divides the exchange (x) contributions into short range (SR) and long range (LR)\cite{HSE_the_paper}.}
  \label{tbl:gksrecap}
  \renewcommand{\arraystretch}{1.5}
  \begin{tabular}{c|ccc}
    \hline\hline
     functional &  semilocal exchange & nonlocal exchange  & correlation \\
    \hline
    PBE\textsuperscript{\emph{a}} & $V_x^{PBE}(\vc{r})$ & - & $V_c^{PBE}(\vc{r})$ \\
    PBE0\textsuperscript{\emph{b}} & $0.75\,V_x^{PBE}(\vc{r})$ & $0.25\,V_x^{EXX}(\vc{r},\vc{r}^\prime)$ & $V_c^{PBE}(\vc{r})$ \\
    EXXc\textsuperscript{\emph{c}} & - & $V_x^{EXX}(\vc{r},\vc{r}^\prime)$ & $V_c^{PBE}(\vc{r})$ \\
    B3LYP\textsuperscript{\emph{d}} & $0.08\,V_x^{LDA}(\vc{r})+ 0.72\,V_x^{PBE}(\vc{r})$ & $0.2\,V_x^{EXX}(\vc{r},\vc{r}^\prime)$ & $0.19\,V_c^{LDA}(\vc{r})+0.81\,V_c^{PBE}(\vc{r})$ \\
    HSE\textsuperscript{\emph{e}} & $0.75\,V_x^{PBE,SR}(\vc{r};\mu)+ \,V_x^{PBE,LR}(\vc{r};\mu)$ & $0.25\,V_x^{EXX,SR}(\vc{r},\vc{r}^\prime;\mu)$ & $V_c^{PBE}(\vc{r})$  \\
    \hline\hline
  \end{tabular} \\
  \textsuperscript{\emph{a}} Ref.~[\onlinecite{PBE_the_paper}], 
  \textsuperscript{\emph{b}} Ref.~[\onlinecite{PBE0_the_paper}], 
  \textsuperscript{\emph{c}} Ref.~[\onlinecite{Aulbur_2000_GWpaper}], 
  \textsuperscript{\emph{d}} Ref.~[\onlinecite{B3LYP_the_paper}], 
  \textsuperscript{\emph{e}} Ref.~[\onlinecite{HSE_the_paper}]
\end{table}
%

\begin{table}
\footnotesize
  \caption{Vertical ionization potential (VIP, eV) of closed shell molecules. Experimental values are taken from the NIST computational chemistry database\cite{nist}. Each column reports the VIP obtained with the \texttt{West} code by performing $G_0W_0$ calculations starting from the solutions of the Kohn-Sham equations with the exchange and correlation potential (see Tab.~\ref{tbl:gksrecap}), specified within parentheses on the first row. In parentheses we report the absolute value of the HOMO energy prior to the application of $G_0W_0$ corrections. ME, MAE, MRE and MARE stand for mean, mean absolute, mean relative and mean relative absolute error as compared to the experiment, respectively.}
  \label{tbl:vipGW}
  \begin{tabular}{l|ccccc|c}
    \hline\hline
    Molecule  & $G_0W_0$(PBE) & $G_0W_0$(PBE0) & $G_0W_0$(EXXc) & $G_0W_0$(B3LYP) & $G_0W_0$(HSE) & Exp. \\
    \hline
    \ce{C_2H_2} 	& 11.10	(7.20) &	 11.38	(8.43)  &   	11.66	(12.19) &	11.37	(8.45) &		11.30 	(8.03) & 11.49 \\
    \ce{C_2H_4} 	& 10.35	(6.74) &	 10.56	(7.86)  &	10.74	(11.26) &	10.58	(7.88) &		10.50	(7.46) & 10.68 \\
    \ce{C_4H_4S}	& 8.90	(5.98) &	 9.15	(7.01)  &	9.44 	(10.12)	& 	9.16 	(7.05) &		9.08	 	(6.62) &  8.86 \\
    \ce{C_6H_6} 	&  9.10	(6.33) &	 9.32	(7.30)  &	9.54		(10.21) &	9.34 	(7.33) &		9.25		(6.91) &  9.25 \\
    \ce{CH_3Cl} 	& 11.27	(7.10) &	 11.57	(8.50)  &	11.89	(12.81) &	11.56	(8.57) &		11.49	(8.11) & 11.29 \\
    \ce{CH_3OH} 	& 10.47	(6.24) &	 10.93	(7.91)  &	11.52	(13.05) &	10.86	(8.01) &		10.82	(7.51) & 10.96 \\
    \ce{CH_3SH} 	& 9.31	(5.55) &	 9.57	(6.78)  &	9.83		(10.58) &	9.59 	(6.87) &		9.49 	(6.39) &  9.44 \\
    \ce{CH_4} 	& 13.99	(9.46) &	 14.34	(10.99) &	14.78	(15.71) &	14.34 	(11.08)&		14.26	(10.60) & 14.40 \\
    \ce{Cl_2} 	& 11.48	(7.28) &	 11.78	(8.69)  &	12.14	(13.02) &	11.77	(8.77) &		11.70	(8.29) & 11.49 \\
    \ce{ClF} 	& 12.47	(7.83) &	 12.84	(9.44)  &	13.35	(14.33) &	12.83	(9.53) &		12.76	(8.04) & 12.77 \\
    \ce{CO} 		& 13.45	(9.06) &	 14.01	(10.74) &	14.88	(15.91) &	13.99	(10.88) &	13.91	(10.34) & 14.01 \\
    \ce{CO_2} 	& 13.31	(9.08) &	 13.73	(10.69) &	14.34	(15.77) &	13.65	(10.76) &	13.64	(10.29) & 13.78 \\
    \ce{CS} 		& 10.92	(7.38) &	 11.53	(8.89)  &	12.51	(13.55) &	11.49	(9.02) &		11.40	(8.49) & 11.33\textsuperscript{\emph{a}} \\
    \ce{F_2} 	& 14.90	(9.42) &	 15.51	(11.73) &	16.34	(18.87) &	15.42	(11.82) &	15.40	(11.33) & 15.70 \\
    \ce{H_2CO} 	& 10.38	(6.25) &	 10.85	(7.84)  &	11.43	(12.82) &	10.78	(7.97) &		10.74	(7.44) & 10.89 \\
    \ce{H_2O} 	& 11.81	(7.23) &	 12.37	(9.04)  &	12.91	(14.67) &	12.31	(9.12) &		12.24	(8.63) & 12.62\textsuperscript{\emph{a}} \\
    \ce{H_2O_2} 	& 10.96	(6.43) &	 11.47	(8.29)  &	12.13	(14.06) &	11.40	(8.40) &		11.36	(7.88) & 11.70 \\
    \ce{HCl} 	& 12.54	(8.03) &	 12.84	(9.48)  &	13.12	(13.93) &	12.85	(9.54) &		12.76	(9.08) & 12.74\textsuperscript{\emph{a}} \\
    \ce{HCN} 	& 13.30	(9.02) &	 13.63	(10.39) &	13.96	(14.54) &	13.60	(10.40) &	13.55	(9.99) & 13.71 \\
    \ce{HF} 		& 15.14	(9.64) &	 15.72	(11.80) &	16.28	(18.52) &	15.65	(11.86) &	15.60	(11.39) & 16.12 \\
    \ce{HOCl} 	& 10.93	(6.61) &	 11.32	(8.18)  &	11.84	(12.97) &	11.28	(8.29) &		11.23	(7.79) & 11.12\textsuperscript{\emph{a}} \\
    \ce{Li_2} 	& 5.03	(3.23) &	 5.29	(3.80)  &	5.39	 	(5.55) &		5.29		(3.87) &		5.14		(3.43) &  5.11\textsuperscript{\emph{a}} \\
    \ce{LiF} 	& 9.97	(6.07) &	 10.85	(7.88)  &	11.45	(13.74) &	10.79	(7.95) &		10.66	(7.47) & 11.30\textsuperscript{\emph{a}} \\
    \ce{LiH} 	& 6.58	(4.35) &	 7.57	(5.42)  &	8.29		(8.86) &	 	7.51		(5.53) &		7.26 	(5.02) &  7.90\textsuperscript{\emph{a}} \\
    \ce{N_2} 	& 14.95	(10.29)& 15.54	(12.20) &	17.23	(17.80) &	15.48	(12.34) &	15.43	(11.80) & 15.58 \\    
    \ce{N_2H_4} 	& 9.28	(5.28) &	 9.72	(6.80)  &	10.24	(11.55) &	9.68		(6.92) &		9.62		(6.40) &  8.98 \\
    \ce{Na_2} 	& 4.73	(3.13) &	 4.86	(3.60)  &	4.88		(5.04)	 &  	4.89		(3.72) &		4.78		(3.24) &  4.89\textsuperscript{\emph{a}} \\
    \ce{NaCl} 	& 8.30	(5.23) &	 9.12	(6.48)  &	9.49		(10.47) &	9.09		(6.53) &		8.93	  	(6.08) &  9.80 \\
    \ce{NH_3} 	& 10.20	(6.16) &	 10.72	(7.71)  &	11.26	(12.55) &	10.68	(7.80) &		10.59	(7.31) & 10.82 \\
    \ce{P_2} 	& 10.44	(7.11) &	 10.62	(8.09)  &	10.76	(11.06) &	10.63	(8.12) &		10.56	(7.70) & 10.62 \\
    \ce{PH_3} 	& 10.46	(6.73) &	 10.70	(7.88)  &	10.94	(11.45) &	10.73	(7.99) &		10.63	(7.49) & 10.59 \\
    \ce{SH_2} 	& 10.26	(6.29) &	 10.53	(7.55)  &	10.76	(11.40) &	10.55	(7.62) &		10.45	(7.15) & 10.50 \\
    \ce{Si_2H_6}	& 10.45	(7.18) &	 10.71	(8.28)  &	11.06	(11.75) &	10.78	(8.40) &		10.64	(7.90) & 10.53 \\
    \ce{SiH_4} 	& 12.44	(8.52) &	 12.82	(9.86)  &	13.32	(14.03) &	12.83	(9.97) &		12.72	(9.46) & 12.30 \\
    \ce{SiO} 	& 11.09	(7.49) &	 11.51	(8.84)  &	12.10	(12.75) &	11.47	(8.94) &		11.41	(8.44) & 11.49 \\
    \ce{SO_2} 	& 11.96	(8.08) &	 12.44	(9.61)  &	13.15	(14.39) &	12.39	(9.75) &		12.34	(9.22) & 12.50 \\
    \hline
    ME (\text{eV}) 	& -0.42	(-4.29) &	0.00	(-2.87)	& 0.49	 (1.50)	& -0.02	(-2.78) &	-0.10	(-3.27) & -- \\
    MAE (\text{eV}) 	& 0.44	(4.29)	& 0.19	(2.87) 	&	  0.51 	(1.50) 	& 0.21 	(2.78) &  	0.22	 (3.27) & -- \\
    MRE(\%) 	 	    	& -3.68	(-37.98) 	& 0.15	 (-25.51)	&	4.31 	(13.00)	& -0.02	  (-24.70) & 	 -0.86	(-29.22) & -- \\
    MARE(\%) 		& 3.96	(37.98) 	& 1.78 	(25.51) &	4.49 	(13.00) 	& 1.96 	(24.70) 	& 2.03 	(29.22) & -- \\
    \hline\hline
  \end{tabular}\\
  \textsuperscript{\emph{a}} The NIST computational chemistry database\cite{nist} does not report the VIP but the ionization potential.
\end{table}

%
%
%
%
%

\begin{table}
\footnotesize
  \caption{Vertical ionization potential (VIP, eV) of open shell molecules. Experimental values are taken from the NIST computational chemistry database\cite{nist}. Each column reports the VIP obtained with the \texttt{West} code by performing $G_0W_0$ calculations starting from the solutions of the Kohn-Sham equations with the exchange and correlation potential (LDA or PBE), specified within parentheses on the first row. In parentheses we report the absolute value of the HOMO energy prior to the application of $G_0W_0$ corrections.}
  \label{tbl:openshellmol}
  \begin{tabular}{l|c|cc|c}
    \hline\hline
    Molecule  & spin & $G_0W_0$(LDA) & $G_0W_0$(PBE)  & Exp. \\
    \hline
    \ce{CF} 	    & 0.5          &	 8.92	(4.68)  &   8.69	(4.72)  &   9.55 \\
    \ce{NF} 	    & 1.0          &	 12.18	(7.14)  &   11.81	(7.05)  &   12.63 \\
    \ce{NO_2} 	& 0.5          &	 10.82	(6.63)  &   10.46	(6.55)  &   11.23 \\
    \ce{O_2} 	& 1.0          &	 12.11	(6.92)  &   11.67	(6.87)  &   12.33 \\
    \ce{S_2} 	& 1.0          &	 9.53	(5.86)  &   9.34	(5.82)  &   9.55 \\
    \hline\hline
  \end{tabular}
\end{table}

\clearpage

\subsection{Electronic structure of crystalline, amorphous and aqueous systems}
\label{sec:VIPcond}

\begin{figure}
\includegraphics[width=0.48\textwidth]{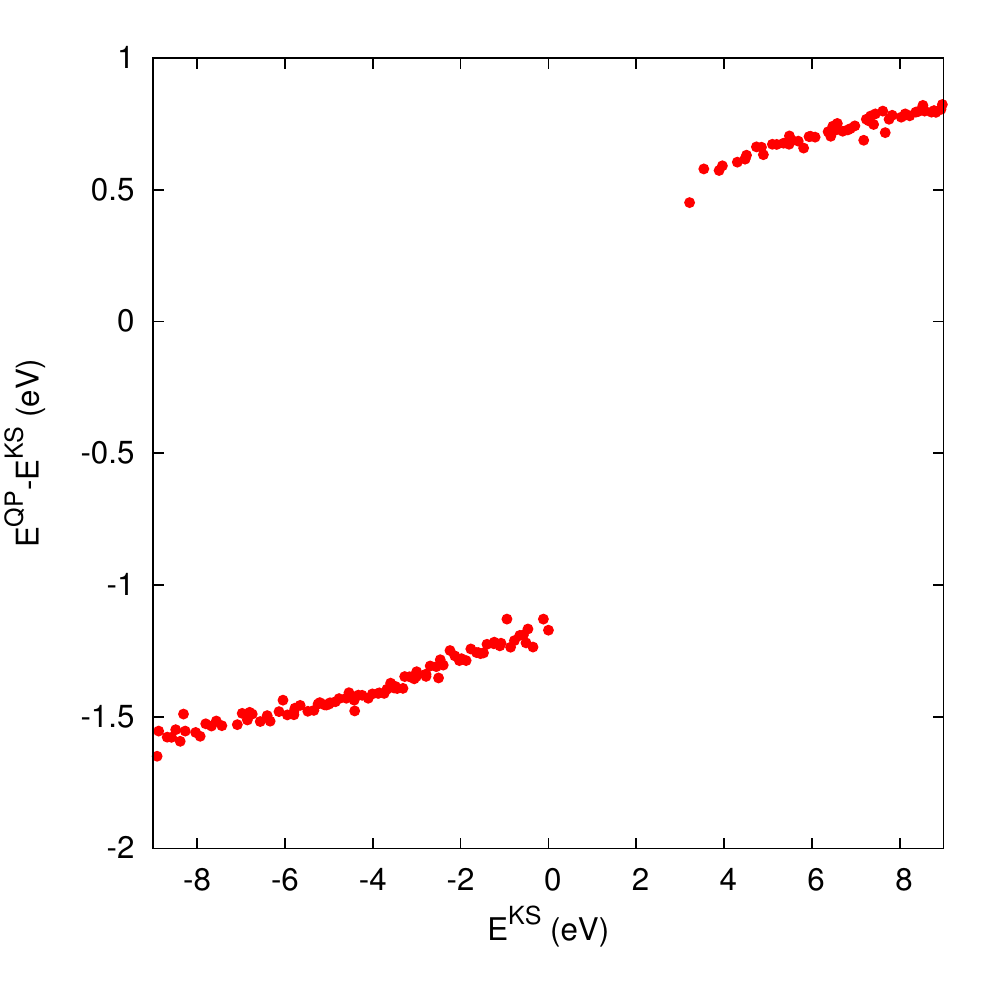}
\includegraphics[width=0.48\textwidth]{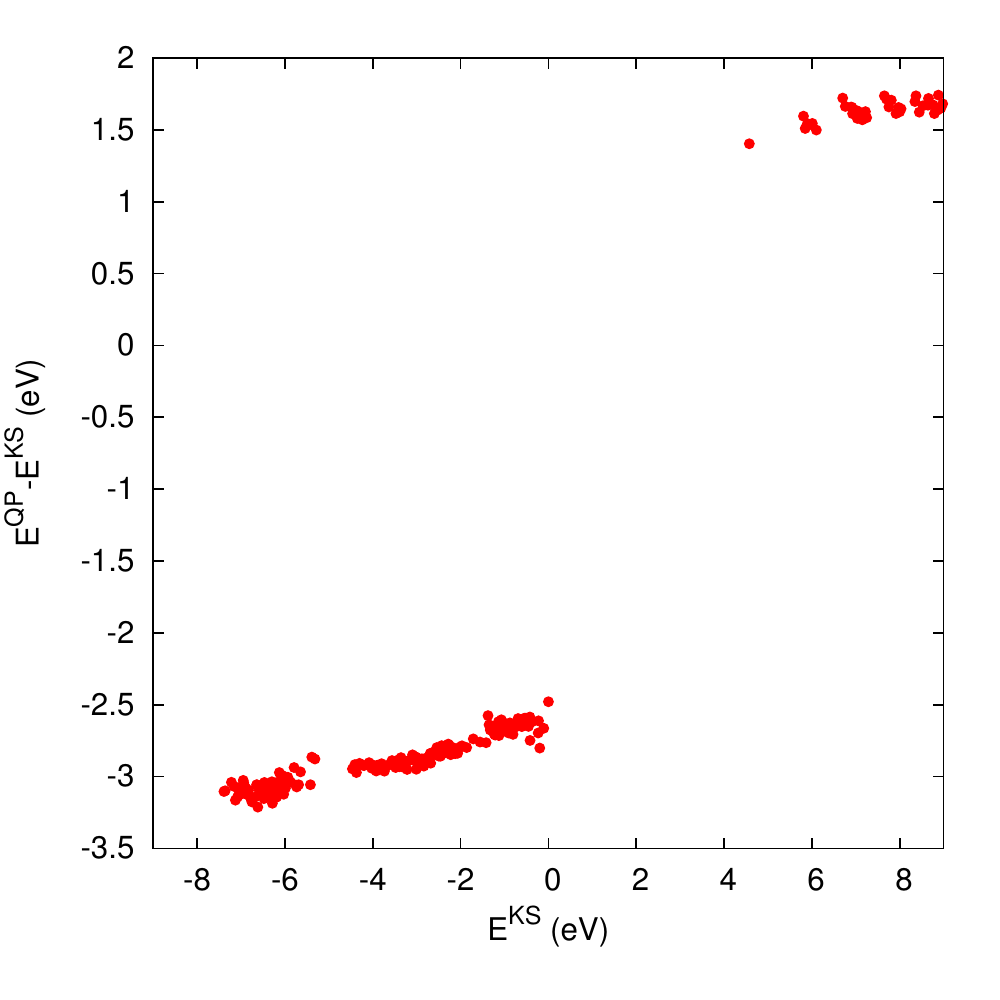}
\caption{(Color online) Quasiparticle (QP) corrections for a configuration of amorphous \ce{Si_3N_4} (left panel) and liquid water at ambient conditions (right panel).}
  \label{fgr:amo}
\end{figure}

We considered three crystalline systems \ce{Si}, \ce{SiC} and \ce{AlAs}, one amorphous \ce{Si_3N_4}, and one liquid water snapshot. The KS electronic structure was computed using super cells and the $\Gamma$ point: $64$ atoms and $256$ valence electrons for \ce{Si}, \ce{SiC} and \ce{AlAs}, with cell edges of $20.53$, $16.48$ and $21.34\,\text{bohr}$, respectively; the amorphous \ce{Si_3N_4} sample consisted of $56$ atoms and $256$ electrons. For \ce{Si}, \ce{SiC}, \ce{AlAs} and amorphous \ce{Si_3N_4} we used a kinetic energy cutoff of $60\,\text{Ry}$. The snapshot of 64 water molecules is taken from a 20~ps trajectory of a Born-Oppenheimer \textit{ab initio} molecular dynamics simulation of liquid water (see Ref.~[\onlinecite{Anh_2014_waterGW}]); and it was described with a cutoff of $85\,\text{Ry}$. In our $G_0W_0$ calculations for condensed systems we used $N_{pdep}=1024$.\\
The QP energies of the crystalline solids at high symmetry k-points are reported in Table~\ref{tbl:SiQP},~\ref{tbl:SiCQP} and~\ref{tbl:AlAsQP}, where  KS energies are given within brackets. The results obtained with \texttt{West} compare well with those of other  plane wave pseudopotential calculations and with experiments.\\
The QP corrections of amorphous \ce{Si_3N_4} and liquid water are reported in Fig.~\ref{fgr:amo}, where again we found that the \texttt{West} results compare well with those of existing calculations\cite{Anh_2014_waterGW,Garbuio_2006_waterGWBSE} and experiments\cite{Bauer_1977_amorphousSi3N4}.

\begin{table}
\footnotesize
  \caption{Quasiparticle (QP) energies of \ce{Si} at high symmetry points, compared with previous calculations and experiment.}
  \label{tbl:SiQP}
  \begin{tabular}{c|cc|c|c}
    \hline\hline
     k-point &  $G_0W_0$(LDA) & $G_0W_0$(PBE) & Theo. & Exp. \\
    \hline
    $L_{1c}$ 	 	   			& 2.26 (1.47) 	& 2.29 (1.59) 	& 2.21\textsuperscript{\emph{a}}, 2.14\textsuperscript{\emph{c}}, 2.18\textsuperscript{\emph{d}}, 2.13\textsuperscript{\emph{e}}, 2.19\textsuperscript{\emph{f}}, 2.05\textsuperscript{\emph{g}} & 2.1\textsuperscript{\emph{j}}, 2.4$\pm$0.1\textsuperscript{\emph{k}} \\
    $L^\prime_{3v}$  			& -1.25 (-1.20) 	& -1.21 (-1.20) 	& -1.23\textsuperscript{\emph{a}}, -1.17\textsuperscript{\emph{c}}, -1.20\textsuperscript{\emph{d}}, -1.25\textsuperscript{\emph{e}}, -1.25\textsuperscript{\emph{f}}, -1.16\textsuperscript{\emph{g}} & -1.2$\pm$0.2\textsuperscript{\emph{h}} \\
    							&				&				&		& 	\\
    $\Gamma_{15c}$   			& 3.35 (2.54) 	& 3.32 (2.55)	& 3.25\textsuperscript{\emph{a}}, 3.24\textsuperscript{\emph{b}}, 3.24\textsuperscript{\emph{c}}, 3.23\textsuperscript{\emph{d}}, 3.25\textsuperscript{\emph{e}}, 3.36\textsuperscript{\emph{f}}, 3.09\textsuperscript{\emph{g}} & 3.40\textsuperscript{\emph{h}}, 3.05\textsuperscript{\emph{i}} \\
    $\Gamma^\prime_{25v}$   	& 0.0 (0.0) 		& 0.0 (0.0)		& 0.0    & 0.0  \\
    							&				&				&		& 		\\
    $X_{1c}$ 				  	& 1.44 (0.63) 	& 1.37 (0.72)	& 1.36\textsuperscript{\emph{a}}, 1.41\textsuperscript{\emph{b}}, 1.34\textsuperscript{\emph{c}}, 1.35\textsuperscript{\emph{d}}, 1.31\textsuperscript{\emph{e}}, 1.43\textsuperscript{\emph{f}}, 1.01\textsuperscript{\emph{g}} & 1.3\textsuperscript{\emph{h}}, 1.25\textsuperscript{\emph{i}} \\
    $X_{4v}$ 				  	& -2.92 (-2.87) 	& -2.96 (-2.85)	& -2.88\textsuperscript{\emph{a}}, -2.80\textsuperscript{\emph{b}}, -2.80\textsuperscript{\emph{c}}, -2.83\textsuperscript{\emph{d}}, -2.96\textsuperscript{\emph{e}}, -2.93\textsuperscript{\emph{f}}, -2.90\textsuperscript{\emph{g}} & -2.90\textsuperscript{\emph{l}}, -3.3$\pm$0.2\textsuperscript{\emph{m}} \\
    \hline\hline
  \end{tabular}\\
  \textsuperscript{\emph{a}} Ref.~[\onlinecite{Anh_2013_GW}], 
  \textsuperscript{\emph{b}} Ref.~[\onlinecite{Umari_2009_GWW}], 
  \textsuperscript{\emph{c}} Ref.~[\onlinecite{Rieger_1999_GW}], 
  \textsuperscript{\emph{d}} Ref.~[\onlinecite{Fleszar_1997_GW}], 
  \textsuperscript{\emph{e}} Ref.~[\onlinecite{Aulbur_2000_GWpaper}], 
  \textsuperscript{\emph{f}} Ref.~[\onlinecite{Rohlfing_1993_gaussGW}],
  \textsuperscript{\emph{g}} Ref.~[\onlinecite{Labegue_2003_GWpaw}], \\ 
  \textsuperscript{\emph{h}} Ref.~[\onlinecite{landolt_1967_book}], 
  \textsuperscript{\emph{i}} Ref.~[\onlinecite{Ortega_1993_inversePE}], 
  \textsuperscript{\emph{j}} Ref.~[\onlinecite{Hulthen_1976_silicon}], 
  \textsuperscript{\emph{k}} Ref.~[\onlinecite{Straub_1985_siliconIPE}], 
  \textsuperscript{\emph{l}} Ref.~[\onlinecite{Spicer_1968_book}], 
  \textsuperscript{\emph{m}} Ref.~[\onlinecite{Wachs_1985_ARPESSi}]
\end{table}

\begin{table}
\footnotesize
  \caption{Quasiparticle (QP) energies of \ce{SiC} at high symmetry points, compared with previous calculations and experiment.}
  \label{tbl:SiCQP}
  \begin{tabular}{c|cc|c|c}
    \hline\hline
     k-point &  $G_0W_0$(LDA) & $G_0W_0$(PBE) & Theo. & Exp. \\
    \hline
    $L_{1c}$ 	 	   			& 6.46 (5.15) & 6.37 (5.19)	& 6.43\textsuperscript{\emph{a}}, 6.30\textsuperscript{\emph{b}}, 6.45\textsuperscript{\emph{c}} 	& 6.35\textsuperscript{\emph{d}} \\
    $L_{3v}$ 		 			& -1.18 (-1.09) & -1.16 (-1.06) 	& -1.10\textsuperscript{\emph{a}}, -1.21\textsuperscript{\emph{b}} & -1.15\textsuperscript{\emph{d}}  \\
    							&				&		& \\
    $\Gamma_{1c}$   			& 7.42 (6.29) & 7.52 (6.29)	& 7.26\textsuperscript{\emph{a}}, 7.19\textsuperscript{\emph{b}}, 7.23\textsuperscript{\emph{c}} 	& 7.4\textsuperscript{\emph{e}} \\
    $\Gamma^\prime_{15v}$   	& 0.0 (0.0) 	& 0.0 (0.0)	& 0.0 	& 0.0 \\
    							&				&		& 	\\
    $X_{1c}$ 				  	& 2.45 (1.29) & 2.28 (1.35) 	& 2.31\textsuperscript{\emph{a}}, 2.19\textsuperscript{\emph{b}}, 1.80\textsuperscript{\emph{c}} 	& 2.39\textsuperscript{\emph{d}}, 2.42\textsuperscript{\emph{d}} \\
    $X_{5v}$ 				  	& -3.46 (-3.25) & -3.46 (-3.19) 	& -3.47\textsuperscript{\emph{a}}, -3.53\textsuperscript{\emph{b}} & -3.6\textsuperscript{\emph{d}} \\
    \hline\hline
  \end{tabular}\\
  \textsuperscript{\emph{a}} Ref.~[\onlinecite{Anh_2013_GW}],
  \textsuperscript{\emph{b}} Ref.~[\onlinecite{Fleszar_1997_GW}],
  \textsuperscript{\emph{c}} Ref.~[\onlinecite{Labegue_2003_GWpaw}],
  \textsuperscript{\emph{d}} Ref.~[\onlinecite{landolt_1967_book}], 
  \textsuperscript{\emph{e}} Ref.~[\onlinecite{Lambrecht_1994_SiC}]
\end{table}

\begin{table}
\footnotesize
  \caption{Quasiparticle (QP) energies of \ce{AlAs} at high symmetry points, compared with previous calculations and experiment.}
  \label{tbl:AlAsQP}
  \begin{tabular}{c|cc|c|c}
    \hline\hline
     k-point &  $G_0W_0$(LDA) & $G_0W_0$(PBE) & Theo. & Exp. \\
    \hline
    $L_{1c}$ 	 	   			& 3.08 (2.15) & 2.94 (2.15) 	& 3.02\textsuperscript{\emph{a}}, 2.84\textsuperscript{\emph{b}}, 2.99\textsuperscript{\emph{c}} 	& 2.36\textsuperscript{\emph{e}} \\
    $L_{3v}$ 		 			& -0.86 (-0.80) & -0.90 (-0.84)	& -0.9\textsuperscript{\emph{a}}, -0.87\textsuperscript{\emph{b}} 		& -  \\
    							&				&		 		& \\
    $\Gamma_{1c}$   			& 3.15 (2.20) & 2.99 (2.23)	& 2.96\textsuperscript{\emph{a}}, 2.74\textsuperscript{\emph{b}}, 2.72\textsuperscript{\emph{c}} 	& 3.13\textsuperscript{\emph{d}} \\
    $\Gamma^\prime_{15v}$   	& 0.0 (0.0) 	& 0.0 (0.0)	& 0.0 		& 0.0 \\
    							&				&				& 	\\
    $X_{1c}$ 				  	& 2.20 (1.35) & 2.01 (1.37)	& 2.13\textsuperscript{\emph{a}}, 2.16\textsuperscript{\emph{b}}, 1.57\textsuperscript{\emph{c}} 	& 2.23\textsuperscript{\emph{d}} \\
    $X_{5v}$ 				  	& -2.25 (-2.15) & -2.35 (-2.21) 	& -2.20\textsuperscript{\emph{a}}, -2.27\textsuperscript{\emph{b}}     & -2.41\textsuperscript{\emph{d}} \\
    \hline\hline
  \end{tabular}\\
  \textsuperscript{\emph{b}} Ref.~[\onlinecite{Anh_2013_GW}],
  \textsuperscript{\emph{b}} Ref.~[\onlinecite{Fleszar_1997_GW}],
  \textsuperscript{\emph{b}} Ref.~[\onlinecite{Labegue_2003_GWpaw}],
  \textsuperscript{\emph{b}} Ref.~[\onlinecite{landolt_1967_book}],
  \textsuperscript{\emph{a}} Ref.~[\onlinecite{Lee_1980_AlAs}]
\end{table}

\clearpage

\section{Large scale calculations}
\label{sec:largesystems}

The method discussed in Sec.~\ref{sec:methodintro}, implemented in the \texttt{West} code and validated in Sec.~\ref{sec:validation} may be used to perform highly parallel $G_0W_0$ calculations and tackle large systems, with $>1000$ of valence electrons in the unit cell. We discuss the performance of the method for both finite and periodic systems, in particular for Si nanocrystals and interfaces of functionalized Si surfaces and water, with up to 1344 and 1560 valence electrons in the unit cell, respectively.

\subsection{Silicon nanocrystals}
\label{sec:sincs}

\begin{figure}
\includegraphics[width=0.68\textwidth]{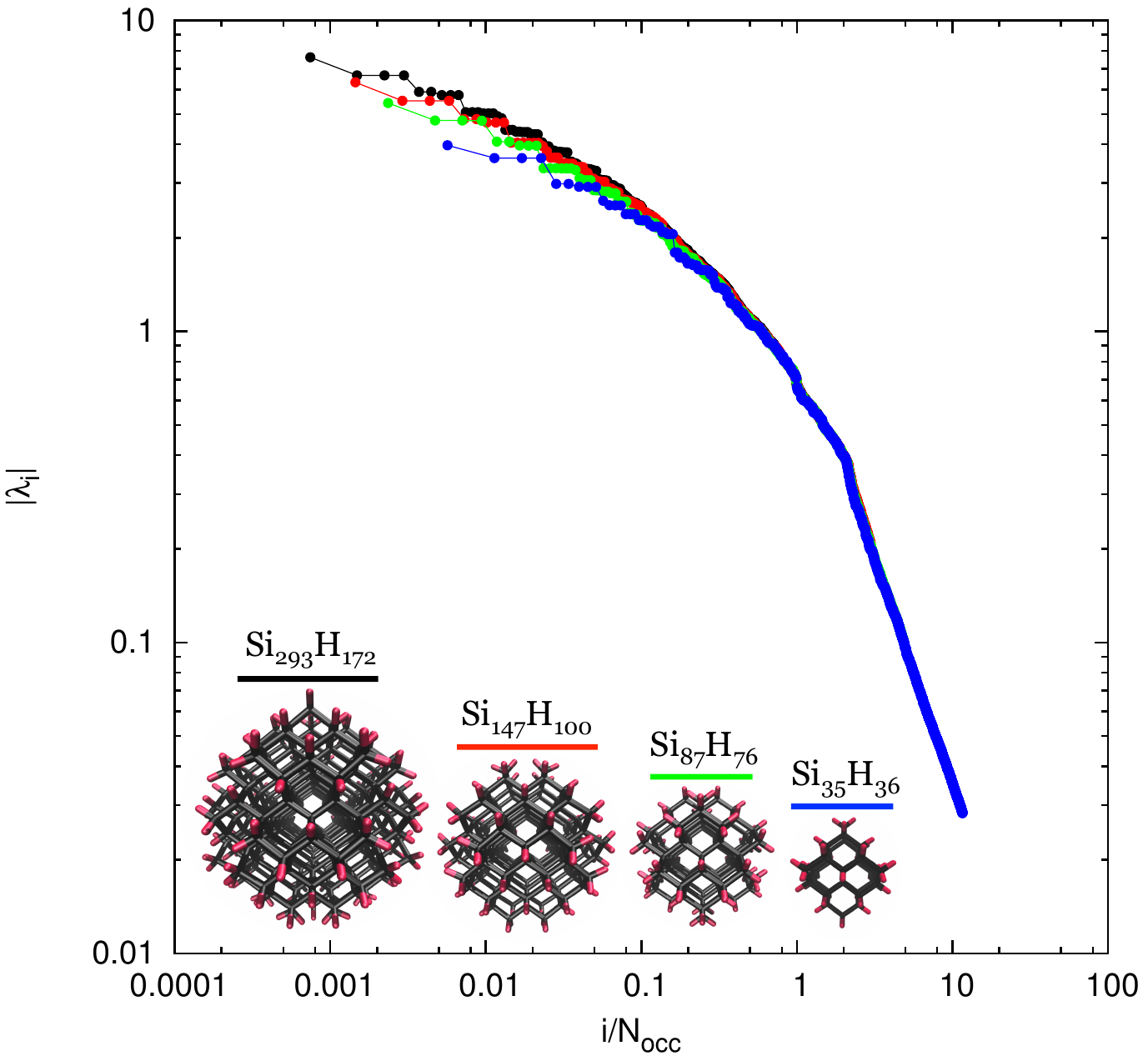}
  \caption{(Color online) PDEP eigenvalues ($\lambda_i$) for the considered Si-NCs.}
  \label{fgr:si-ncs-pdep}
\end{figure}

We considered four Si-NCs: \ce{Si_{35}H_{36}} ($1.3\,\text{nm}$), \ce{Si_{87}H_{76}} ($1.6\,\text{nm}$), \ce{Si_{147}H_{100}} ($1.9\,\text{nm}$), and \ce{Si_{293}H_{172}} ($2.4\,\text{nm}$)\bibnote{We have indicated in parentheses the diameter of each Si nanocrystal.}. The structure of each NCs was obtained by carving out of bulk Si a sphere of Si atoms of given radius, by terminating all dangling bonds with H atoms and by relaxing the NC structure within DFT-PBE. A kinetic energy cutoff of $25\,\text{Ry}$, PBE norm-conserving pseudopotentials and a cubic cell of edge $90\,\text{bohr}$ were used. The computed HOMO and LUMO energies and the energy gap ($\text{E}_{\text{gap}}$) are reported in Table~\ref{tbl:si-ncs}. The HOMO and LUMO energies referred to vacuum were obtained using the Makov-Payne\cite{makov_1995} method. For each Si-NCs we considered $N_{pdep}=2048$. PDEP eigenvalues are reported in Fig.~\ref{fgr:si-ncs-pdep} and they clearly show that the only difference between Si-NCs of different size appears for the most screened eigenpotentials. As discussed in Sec.~\ref{sec:pdepmethod}, the PDEP eigenvalues of the least screened eigenpotentials are weakly affected by the microscopic structure of the system and may likely be predicted by model screening functions. The computed $G_0W_0$ energy gaps for \ce{Si_{35}H_{36}}, \ce{Si_{87}H_{76}}, \ce{Si_{147}H_{100}} and \ce{Si_{293}H_{172}} are $6.29$, $4.77$, $4.20$ and $3.46\,\text{eV}$, respectively. These results are in good agreement with those of other recent calculations using MBPT or $\Delta$SCF method.\cite{baer_2014_stGW}, although our computed HOMO and LUMO energies differ slightly from those reported in Ref.~[\onlinecite{baer_2014_stGW}].

\begin{table}
\small
  \caption{Quasiparticle (QP) energies and energy gap ($\text{E}_{\text{gap}}$) of Si nanocrystals. The Kohn-Sham eigenvalues obtained using the PBE exchange-correlation functional are reported in parentheses.}
  \label{tbl:si-ncs}
  \begin{tabular}{lcccc}
    \hline\hline
    \multirow{2}{*}{Si-NC}  & \multirow{2}{*}{$N_{occ}$} & HOMO (eV) & LUMO (eV) & $\text{E}_{\text{gap}}$ (eV) \\\cmidrule{3-5}
                         &      &  $G_0W_0$ (PBE) & $G_0W_0$ (PBE) &  $G_0W_0$ (PBE) \\\hline
    \ce{Si_{35}H_{36}} 	 & 176  & -7.59 (-6.08) & -1.30 (-2.57) & 6.29 (3.51)  \\
    \ce{Si_{87}H_{76}} 	 & 424  & -6.69 (-5.54) & -1.92 (-2.96) & 4.77 (2.58)  \\
    \ce{Si_{147}H_{100}}	 & 688  & -6.48 (-5.44) & -2.27 (-3.15) & 4.21 (2.29)  \\
    \ce{Si_{293}H_{172}}	 & 1344 	& -5.82 (-5.19) & -2.36 (-3.41) & 3.46 (1.78)  \\
    \hline\hline
  \end{tabular}
\end{table}

\clearpage

\subsection{Solid/liquid interfaces}
\label{sec:interfaces}

\begin{figure}
\includegraphics[width=\textwidth]{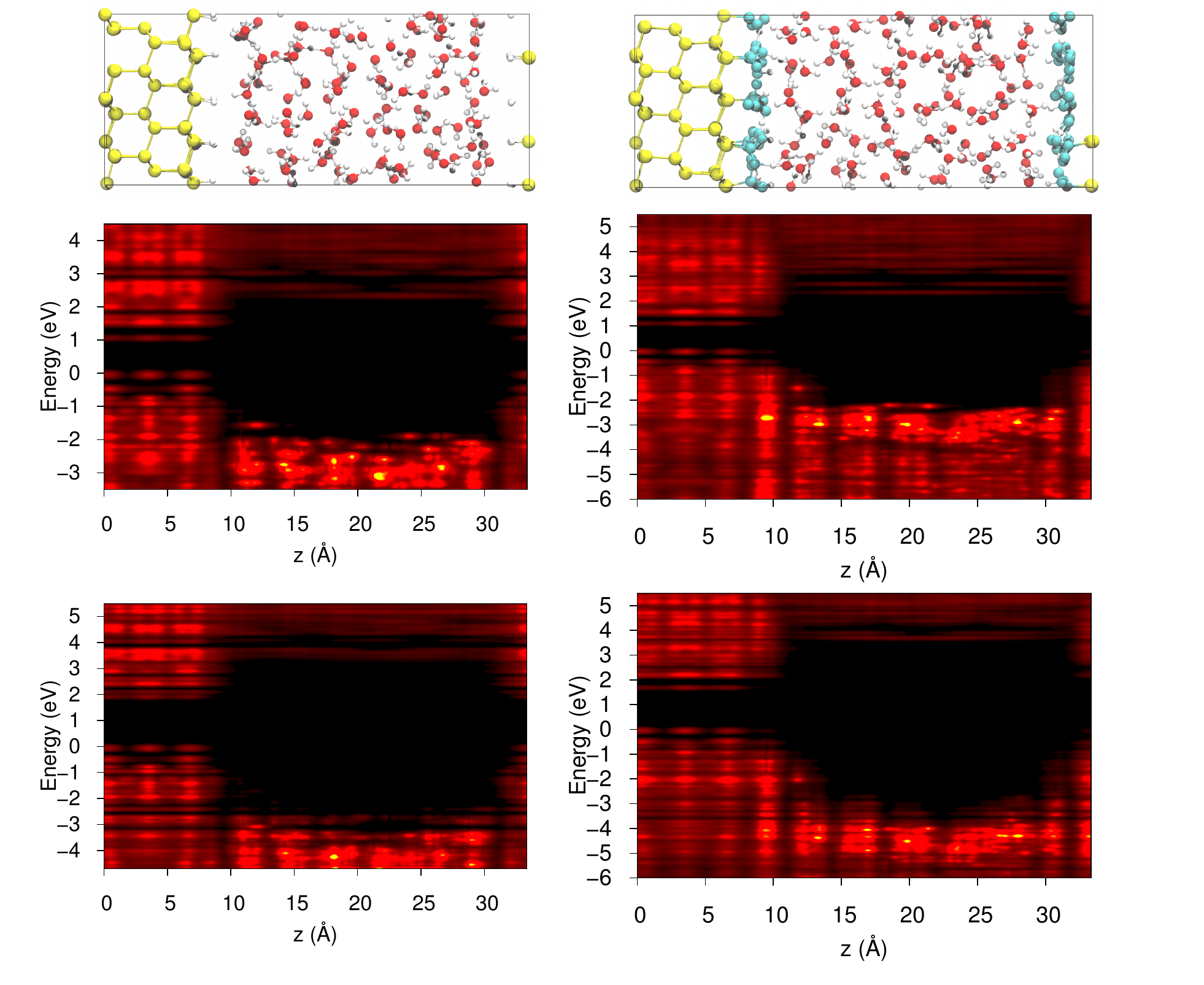}
\caption{(Color online) The local density of states (LDOS, see text) of two solid/liquid interfaces: \ce{H-Si}/\ce{H_2O} (left panels) and \ce{COOH-Si}/\ce{H_2O} (right panels). The top panels report the side view of the unit cells. Bottom (middle) panels report the LDOS obtained using $G_0W_0$ (KS-DFT) energies in Eq.~\eqref{eq:LDOS}. A color scale that ranges from black to red is used to plot the LDOS; black areas indicate energy gap regions.}
\label{fgr:coohinterface}
\end{figure}

We now turn to discuss QP energies of extended, large systems. We considered two solid/liquid interfaces: \ce{H-Si}/\ce{H_2O} and \ce{COOH-Si}/\ce{H_2O}, that were recently studied by T.A.~Pham \textit{et al.}\cite{pham_2014_siwater} to align band edges of functionalized Si(111) surfaces with water reduction and oxidation potentials. The orthorombic unit cell $(L_x\times L_y \times L_z)$ of each system was obtained by interfacing 108 water molecules with 72 Si atoms and by terminating the solid surface exposed to water with 24 \ce{H} atoms or \ce{24} COOH groups, resulting in a $(21.97\times 25.37\times 63.19)\,\text{bohr}^3$ supercell with 1176 valence electrons and a $(21.97\times 25.37\times 67.53)\,\text{bohr}^3$ supercell with 1560 valence electrons for \ce{H-Si}/\ce{H_2O} and for \ce{COOH-Si}/\ce{H_2O}, respectively. Both interface geometries were extracted from a $\sim 30$~ps trajectory of a Car-Parrinello molecular dynamics simulation of the interface where all water molecules and atoms of the semiconductor surfaces were allowed to move (see Ref.~[\onlinecite{pham_2014_siwater}])\bibnote{The comparison between the results reported here and those of Fig. 11 of Ref.~[\onlinecite{Anh_2014_waterGW}] is not straightforward for several reasons: (i) here we just reported results for one snapshot, arbitrarily extracted from a $\sim 30$~ps simulation, as a proof of principle that $G_0W_0$ calculations can indeed be done. (ii) The difference between results obtained by doing $G_0W_0$ calculations of slabs, as opposed to using values of the band edges of water and functionalized Si computed in the bulk, as done in Ref.~[\onlinecite{pham_2014_siwater}], remains to be explored and will be the subject of further investigation in a subsequent work.}. Side views of the unit cells are shown in Fig.~\ref{fgr:coohinterface}, top panels. The KS electronic structure of both systems was obtained at the PBE level of theory using $85\,\text{Ry}$ for the kinetic energy cutoff. The local density of states (LDOS) was obtained from the wavefunctions $\psi_n$ and energy levels $\varepsilon_n$ as
\begin{equation}
\text{LDOS}(z,E) = \sum_n \int\frac{dx}{L_x} \int\frac{dy}{L_y} \left|\psi_n(x,y,z)\right|^2 \delta(E-\varepsilon_n) \label{eq:LDOS}
\end{equation}
where $z$ is the axis perpendicular to the interface and $\delta$ is the Dirac delta function. The LDOS of both systems, obtained at the PBE level of theory, is reported in Fig.~\ref{fgr:coohinterface}, middle panels. Those at $G_0W_0$ level, obtained by replacing the KS energies with QP energies in Eq.~\eqref{eq:LDOS}, are shown in Fig.~\ref{fgr:coohinterface}, bottom panels. The figures show that the method developed in Sec.~\ref{sec:methodintro} can be successfully used to compute the positions of the valence and conduction band edges of a realistic interface and hence to define an electronic thickness of the interface, by analyzing how the bulk eigenvalues are modified in proximity of the interface. The method can of course be used for systems with impurity levels and to investigate semiconductor surfaces interfaced with aqueous solutions containing ions and to study the influence of ions on the electronic structure of the interface. 
The method developed here is not limited to solid/liquid interfaces and to planar geometries and has broad applicability to any complex (nanostructured) materials inclusive of heterogeneous interfaces\cite{giantomassi_2011_interfaces}.

\clearpage

\section{Conclusions}
\label{sec:conclusions}

We presented a formulation of the GW method for large scale calculations carried out with the plane-wave pseudopotential method. The evaluation of polarizabilities and electronic self-energies does not require the explicit computation of virtual states. Polarizabilities and dielectric matrices were represented with a basis set composed of the eigenstates of the dielectric matrix at zero frequency, obtained using iterative procedures. In the calculation of the correlation self-energy we avoided the use of the analytic continuation and carried out the frequency integration by means of a contour deformation technique. In addition we presented a parallel implementation of the method that allowed us to compute the electronic properties of large nanostructures and of solid/liquid interfaces. The method is not restricted to DFT inputs obtained with semi-local functionals but can be used in conjunction with DFT calculations with hybrid functionals.\\
We presented a validation of the method for molecules (open and closed shell) and solids (both crystalline and amorphous) and found good agreement with data previously appeared in the literature for converged calculations. We then applied our technique to silicon nanoparticles (up to a diameter of $2.4\,\text{nm}$) and solid/liquid interfaces (with up to 1560 valence electrons in the unit cell). We showed that it is now possible to carry out many body perturbation theory calculation of realistic slabs representing a semiconductor/water interface and to study in detail the modification of the bulk states at the interfaces and hence define an electronic thickness of the interface. Work is in progress to couple our GW calculations with ab initio molecular dynamics simulations of realistic materials, and to include finite temperature and statistical effects in our MBPT calculations.

\begin{acknowledgement}
This work was supported by the Army Research Laboratory Collaborative Research Alliance in Multiscale Multidisciplinary Modeling of Electronic Materials (CRA-MSME, Grant No. W911NF-12-2-0023) and by DoE grant No. DE-
FG02-06ER46262; the computational resources were provided by DoD Supercomputing Resource Center of the Department of Defense High Performance Computing Modernization Program. An award of computer time was provided by the Innovative and Novel Computational Impact on Theory and Experiment (INCITE) program. This research used resources of the Argonne Leadership Computing Facility at Argonne National Laboratory, which is supported by the Office of Science of the U.S. Department of Energy under contract DE-AC02-06CH11357.
Discussions with T.A. Pham and J.H. Skone are greatly acknowledged. 
We thank B. Rice for her help and support with computational grants of the U.S. Department of Defense's High Performance Computing Modernization Program.
\end{acknowledgement}

%
%

\bibliography{manuscript}

\begin{mcitethebibliography}{127}
\bibitem[Not()]{Note-1}
We use everywhere atomic Rydberg units.\relax
\mciteBstWouldAddEndPunctfalse
\mciteSetBstMidEndSepPunct{\mcitedefaultmidpunct}
{}{\mcitedefaultseppunct}\relax
\EndOfBibitem
\bibitem[Not()]{Note-3}
Although the presented methodology is general, for the purpose of this work we
  have considered only normconserving
  pseudopotentials\cite{focher_1991_pseudoisphysicalion}.\relax
\mciteBstWouldAddEndPunctfalse
\mciteSetBstMidEndSepPunct{\mcitedefaultmidpunct}
{}{\mcitedefaultseppunct}\relax
\EndOfBibitem
\bibitem[Not()]{Note-5}
The divergence of the Coulomb potential present in
  Eq.~\eqref{eq:integrationxiq} can be numerically removed using spherical
  coordinates. The specific shape of the BZ is taken into account by using a
  Monte Carlo integration method.\relax
\mciteBstWouldAddEndPunctfalse
\mciteSetBstMidEndSepPunct{\mcitedefaultmidpunct}
{}{\mcitedefaultseppunct}\relax
\EndOfBibitem
\bibitem[Not()]{Note-6}
In the present formulation we considered integer occupation numbers and only
  the dielectric response given by interband transitions.\relax
\mciteBstWouldAddEndPunctfalse
\mciteSetBstMidEndSepPunct{\mcitedefaultmidpunct}
{}{\mcitedefaultseppunct}\relax
\EndOfBibitem
\bibitem[Not()]{Note-7}
Because of the RPA, Eq.~\eqref{eq:linearalgebrachichi0} formally corresponds to
  the solution of $\bar{\chi}$ when only the Hartree contribution of
  Eq.~\eqref{eq:LFHartree} is considered in Eq.~\eqref{eq:LF}\relax
\mciteBstWouldAddEndPuncttrue
\mciteSetBstMidEndSepPunct{\mcitedefaultmidpunct}
{\mcitedefaultendpunct}{\mcitedefaultseppunct}\relax
\EndOfBibitem
\bibitem[Not()]{Note-9}
We note that the GW technique reported in this work replaces explicit
  summations over empty states present in both $G^\sigma_{KS}$ and $W_p$ with
  projection operations, thus avoiding the calculation of slow converging
  summations over empty states. However, within first order perturbation
  theory, in order to get the QP corrections to the energy of the KS state
  $\psi_{n\vc{k}\sigma}$, one needs to compute the mean value of the
  self-energy operator over that specific state and hence the wave function
  $\psi_{n\vc{k}\sigma}$ needs to be computed. Eq.~\eqref{eq:fcontour} shows
  that only the virtual states with energy $\varepsilon_{m\vc{k}\sigma}$
  between $\varepsilon_F$ and $E^{QP}_{n\vc{k}\sigma}$ are required, which are
  available at no extra cost after $(\varepsilon,\psi)_{n\vc{k}\sigma}$ have
  been obtained from the solution of the KS equations.\relax
\mciteBstWouldAddEndPunctfalse
\mciteSetBstMidEndSepPunct{\mcitedefaultmidpunct}
{}{\mcitedefaultseppunct}\relax
\EndOfBibitem
\bibitem[Not()]{Note-10}
In order to prevent the occurrence of spurious size effects in the computation
  of VIPs, we referred the KS eigenvalues to the vacuum level. The vacuum level
  was computed as the spherical average of the electrostatic potential over a
  sphere of diameter equal to the edge of the cubic simulation cell and
  centered on the ionic pseudo-charge position. Because in our $G_0W_0$
  calculations we did not update the wavefunctions and thus the charge density,
  the vacuum level was not recomputed when QP corrections were added to KS
  eigenvalues.\relax
\mciteBstWouldAddEndPunctfalse
\mciteSetBstMidEndSepPunct{\mcitedefaultmidpunct}
{}{\mcitedefaultseppunct}\relax
\EndOfBibitem
\bibitem[Not()]{Note-11}
For open shell molecules, the VIPs obtained with the LDA (PBE) exchange
  correlation functional have been obtained using LDA (PBE) norm-conserving
  pseudopotentials. Simulations were performed considering collinear spins, a
  cell of edge $30$ bohr and a kinetic energy cutoff of $85$ Ry.\relax
\mciteBstWouldAddEndPunctfalse
\mciteSetBstMidEndSepPunct{\mcitedefaultmidpunct}
{}{\mcitedefaultseppunct}\relax
\EndOfBibitem
\bibitem[Not()]{Note-12}
We have indicated in parentheses the diameter of each Si nanocrystal.\relax
\mciteBstWouldAddEndPunctfalse
\mciteSetBstMidEndSepPunct{\mcitedefaultmidpunct}
{}{\mcitedefaultseppunct}\relax
\EndOfBibitem
\end{mcitethebibliography}

\end{document}